\documentclass[twocolumn,aps,prb,,noeprint]{revtex4-1}

\usepackage[latin9]{inputenc}
\usepackage{amsmath}
\usepackage{amssymb}
\usepackage{cancel}
\usepackage{float}
\usepackage{esint}

\makeatletter
\@ifundefined{textcolor}{}
{%
 \definecolor{BLACK}{gray}{0}
 \definecolor{WHITE}{gray}{1}
 \definecolor{RED}{rgb}{1,0,0}
 \definecolor{GREEN}{rgb}{0,1,0}
 \definecolor{BLUE}{rgb}{0,0,1}
 \definecolor{CYAN}{cmyk}{1,0,0,0}
 \definecolor{MAGENTA}{cmyk}{0,1,0,0}
 \definecolor{YELLOW}{cmyk}{0,0,1,0}
}

\usepackage{graphicx}
\usepackage{epsfig}

\usepackage{amsfonts}
\usepackage{xcolor}

\usepackage{subcaption}
\usepackage[bookmarks=false]{hyperref}
\hypersetup{colorlinks=true, citecolor=blue, urlcolor=blue, linkcolor=blue}

\makeatother

\begin{document}

\title{Current-induced atomic motion, structural instabilities, and negative temperatures on molecule-electrode interfaces in electronic junctions }

\author{Riley J. Preston, Vincent F. Kershaw and Daniel S. Kosov}

\address{College of Science and Engineering, James Cook University, Townsville,
QLD, 4811, Australia }


\begin{abstract}
Molecule-electrode interfaces in molecular electronic junctions are prone to chemical reactions, structural changes, and localized heating effects caused by electric current.  These can be exploited for device functionality or may be degrading processes that limit performance and device lifetime. We develop a nonequilibrium Green's function based transport theory in which the central region atoms and, more importantly, atoms on molecule-electrode interfaces are allowed to move. The separation of time-scales of slow nuclear motion and fast electronic dynamics enables the algebraic solution of the Kadanoff-Baym equations in the Wigner space.  As a result, analytical expressions for dynamical corrections to the adiabatically computed Green's functions are produced.  These dynamical corrections depend not only on the instantaneous molecular geometry but also on the nuclear velocities. To make the theoretical approach fully self-consistent, the same time-separation approach is used to develop expressions for the adiabatic, dissipative, and stochastic components of current-induced forces in terms of adiabatic Green's functions.  Using these current induced forces, the equation of motion for the nuclear degrees of freedom is cast in the form of a Langevin equation.  The theory is applied to model molecular electronic junctions.  We observe that the interplay between the value of the spring constant for the molecule-electrode chemical bond and electronic coupling strength to the corresponding electrode is critical  for the appearance of structural instabilities and, consequently, telegraphic switching in the electric current. The range of model parameters is identified to observe structurally stable molecular junctions as well as various different kinds of current-induced telegraphic switching. The interfacial structural instabilities are also quantified based on current noise calculations.
\end{abstract}
\maketitle

\section{Introduction}

A molecular electronic junction is a single molecule chemically bonded to two macroscopic electrodes. The structural flexibility of the organic framework makes the current-induced atomic motion one of the most critical processes to the performance of molecular electronic devices. Atomic nuclei feel the tug of the tunneling electrons, which can induce nonequilibrium excitations in the molecular vibrations \cite{ryndyk07,ryndyk06,galperin07,galperin06a,hartle08}, atomic rearrangements and rotations \cite{ueba09,ratner13}, as well as
 large-scale current-driven conformational changes such as chemical reactions \cite{ho99,PhysRevLett.85.2777,Repp26052006,ueba09,dzhioev11,catalysis12}, bond ruptures\cite{PhysRevLett.78.4410,peskin18}, telegraphic switching between multiple geometries \cite{noise10,noise14,telegraph16,noise17,kosov18,C4NR03480E,gurvitz16,rudge19-telegraph}, and structural instabilities \cite{PhysRevB.83.035420,pistolesi08, pistolesi12,pistolesi15,pistolesi16,pistolesi18,hector18,pistolesi10}. Current-induced forces exerted by out-of-equilibrium electrons on nuclei result in heating within the system, consequently straining molecular bonds and decreasing the functionality and lifespan of the system. The structural  instabilities and switching between different conformations introduced by these interactions can be detrimental to the performance of a nanoscale system. From another perspective, the small size and high sensitivity of nanoscale junctions result in the capability of utilizing the current-induced forces as a mechanism for satisfying specific tasks: such as a molecular switch, mass and charge sensors, as well as nanoscale motors \cite{EOM2011115}. Moreover, the ability to directly manipulate temperature within the system using quantum mechanical effects such as quantum back-action enables new ways of quantum control of nanomechanical systems \cite{back-action2010}.

One observed consequence of the sensitivity of a molecular junction to the voltage bias is voltage-induced breakage due to quantum heating: molecular junctions can rarely sustain experimentally more than 1-2 V of applied voltage bias \cite{Schulze08,Sabater15}. In molecular junctions, the energy of the flowing current is dispersed by inelastic scattering of electrons through the system. There is a delicate balance between the heating due to inelastic processes and heat dissipation within molecular systems; if a molecular junction is allowed to get too hot, instabilities and bond breakages can occur \cite{hector18}. The absence of the fluctuation-dissipation theorem, which balances these energy gain and loss processes in equilibrium, complicates considerably the theoretical consideration \cite{Bode11,Bode12,catalysis12}.

Another experimental manifestation of the current-induced forces  in molecular junctions is the telegraph noise, which is the stochastic switching over time between two different values of the electric current. Numerous studies of molecular junctions have observed a discrete switching between two or more states in the system, which can be observed via analysis of the measured time-evolution of the current through the system \cite{noise10,noise14,telegraph16,noise17,C4NR03480E}. While the exact source of telegraph noise in molecular junctions depends on the particular experimental configuration, it is usually due to either dynamic switching between two different conformations of the molecular bridge or, more often, due to bond fluctuations of the metal-molecule contact. The telegraph noise is heavily influenced by the presence of inelastic interactions between the tunneling electrons and nuclear degrees of freedom within the junction, which can provide the energy required for conformational changes. In order to ensure the stability of the current-voltage properties of a given molecular junction, we would generally like to avoid these fluctuations between states which may have unforeseen effects. However, perhaps the existence of telegraph noise for specific systems may enable them to replicate certain functions of electric circuits due to their inherent molecular properties \cite{C4NR03480E}.

Many theoretical approaches have been developed to deal with nuclear dynamics in molecular junctions, which generally fall into two categories. The first is based on treating nuclear motion as harmonic vibrations around equilibrium and typically assuming linearized electron-vibration coupling.
Then either a master equation based method \cite{PhysRevB.69.245302,fcblockade05,PhysRevB.83.115414,may02,PhysRevB.94.201407,segal15,kosov17-wtd,kosov17-nonren,dzhioev14,dzhioev15}  or a NEGF method \cite{ryndyk06,dahnovsky:014104,galperin06,ryndyk06,ryndyk07,hartle08,rabani14,hartle15,PhysRevB.75.205413} is used to describe the system.  All  theoretical  methods in the first category assume that the amplitudes of nuclear motion are small and nuclei vibrate harmonically about the zero-current equilibrium geometry. Furthermore, they require that either electron-vibration coupling or molecule-electrode interaction should be small, allowing for a perturbative treatment.

The second approach has gained significant attention recently; it is based on the non-equilibrium Born-Oppenheimer approximation -- nonequilibrium quantum electrons exert non-conservative stochastic forces on  the nuclear degrees of freedom, which are treated classically \cite{pistolesi08,pistolesi10,fuse,dzhioev11,Bode11, Bode12,catalysis12,galperin15,subotnik17,subotnik17-prl,subotnik18,kershaw17,kershaw18,kershaw19}. It does not assume that the amplitude of nuclear motion is small or harmonic, nor is it required that the electron-vibration interaction be treated as small or linear in nuclear displacement. This approach casts the stochastic nuclear dynamics in the form of a Langevin equation. There are a number of varying methods for calculating the friction tensor and random force which are the main ingredients of the Langevin equation. These include the use of NEGF methods \cite{Langreth98,Langreth99,Bode12,PhysRevB.86.195419,subotnik17,subotnik18,Dou2017}, which we will also employ in this paper, as well as scattering theory approaches\cite{Bode12,Bode11,Mozyrsky06,Bennett10}, path integral methods and influence functional methods\cite{Brandbyge1995,Mozyrsky07,Todorov12}.

With a handful of exceptions \cite{subotnik17,subotnik18,Coffman2018,Erpenbeck2019,
Dou2017,peskin18}, all these theoretical approaches largely focus on nuclear motion localized in the central region; however, the motion at the molecule-electrode interface is at least equally important. Large amplitude conformational changes such as chemical reactions, switching between different geometries, localized  heating, and electromigration of atoms predominantly occur on the interface in molecular electronic junctions.
Our goal is to derive a Langevin equation to describe the dynamics of nuclear motion on molecule-electrode interfaces, and then utilize computational simulations to provide insight on the impacts of nuclear motion on the measured current noise through the system with relevance to physical applications. To this end, we do not only obtain the Langevin equation with all parameters fully determined from adiabatic Green's functions, but also solve approximately the time-dependent Kadanoff-Baym equations along the generated stochastic trajectory. The solution of the Kadanoff-Baym equations makes use of Wigner space and gradient expansion methods to separate fast electronic and slow nuclear time-scales. As a result, we have produced a theory where the nuclear motion on molecule-electrode interfaces and electronic dynamics is treated self-consistently.

The paper is organized as follows. Section II describes the theory: solution to the Kadanoff-Baym equations, derivation of dynamical corrections to the current, and also derivation of all components of the current-induced forces in terms of NEGF. The physical model and the results of calculations are presented in section III.  The conclusions of the paper are summarized in section IV. The technical details of electronic diffusion coefficient derivation is relegated to the appendix.

We use atomic units throughout the paper, both in derivations and in the calculations ($\hbar=e=1$).
 
\section{Theory}

\subsection{Hamiltonian}

Let us consider the general tunneling Hamiltonian which describes a
molecular junction 
\begin{equation}
H(t)=H_{M}(t)+H_{L}+H_{R}+H_{LM}(t)+H_{RM}(t).
\label{hamiltonian}
\end{equation}
Here $H_{M}$ is the time-dependent Hamiltonian for the molecule, $H_{L}$ is the Hamiltonian for the left lead and $H_{R}$ is the Hamiltonian for the right lead. The terms $H_{LM}$ and $H_{RM}$ are time-dependent and describe the tunneling of electrons between the molecule and the left and right leads, respectively.

Suppose that  ${x}(t)$ describes the time-dependent trajectory of atomic coordinates including atoms on the molecule-lead interfaces.
We assume that ${x}(t)$ is a classical variable. To simplify the notation, we take $x(t)$ as a scalar rather than a multidimensional vector throughout derivations in the paper, but all our results can be readily extended to the case of many classical variables.

We assume that the molecule contains non-interacting electrons and is described by some quadratic Hamiltonian 
\begin{equation}
H_{M}(t)=\sum_{ij}h_{ij}(x)d_{i}^{\dag}d_{j}.
\end{equation}
Here $d_{i}^{\dag}$ and $d_{j}$ are fermionic creation and annihilation operators for single-particle states localized in the molecular space; $h_{ij}$ is the corresponding matrix elements of the molecular Hamiltonian.
The left and right leads of the molecular junction are macroscopic reservoirs of non-interacting electrons 
\begin{equation}
H_{L} + H_{R}=\sum_{k\alpha} \epsilon_{k\alpha} a_{k\alpha}^{\dag} a_{k\alpha},
\end{equation}
where $a_{k\alpha}^{\dagger}$ creates an electron in the single-particle
state $k$ of the $\alpha=L/R$ (left/right) lead with energy $\epsilon_{k\alpha}$, and $a_{k\alpha}$ is the corresponding electron annihilation operator. The tunneling interaction is 
\begin{equation}
H_{LM}(t)+H_{RM}(t)=\sum_{k\alpha i}v_{k\alpha i}(x)a_{k\alpha}^{\dag}a_{i}+h.c.
\end{equation}
where the tunneling amplitudes $v_{k\alpha i}(x)$ depend on the molecular junction geometry.

The molecular Hamiltonian $H_M(t)$ as well as the tunneling molecule-lead interactions $H_{LM}(t)$ and $H_{RM}(t)$ are explicitly time-dependent via the dependence on the time-evolution of the molecular junction geometry $x(t)$.

\subsection{Non-Adiabatic expansion of Kadanoff-Baym equations in Wigner space}

The basic building blocks in our derivation are non-adiabatic (exact) retarded, advanced and lesser molecular Green's
functions, calculated with a fully time-dependent Hamiltonian along a given trajectory $x(t)$:
\begin{equation}
{\cal G}_{ij}^{R}(t,t')=-i\theta(t-t')\langle\{d_{i}(t),d_{j}^{\dag}(t')\}\rangle,
\end{equation}
\begin{equation}
{\cal G}_{ij}^{A}(t,t')=\Big({\cal G}_{ji}^{R}(t',t)\Big)^{\dag},
\end{equation}
and 
\begin{equation}
{\cal G}_{ij}^{<}(t,t')=i\langle d_{j}^{\dag}(t')d_{i}(t)\rangle.
\end{equation}
These Green's functions are computed using a system of coupled Kadanoff-Baym equations of motion (note that we consider the retarded and advanced equations collectively) \cite{haug-jauho}
\begin{multline}
\Big(i\partial_{t}- h(t)\Big)\mathcal{G}^{R/A}(t,t')=\delta(t-t^{\prime})\\
+\int  dt_{1}\Sigma^{R/A}(t,t_{1})\mathcal{G}^{R/A}(t_{1},t'),\label{eqm2}
\end{multline}
and 
\begin{multline}
\Big(i\partial_{t}- h(t)\Big)\mathcal{G}^{<}(t,t')=\\
\int  dt_{1}\Big(\Sigma^{R}(t,t_{1})\mathcal{G}^{<}(t_{1},t')+\Sigma^{<}(t,t_{1})\mathcal{G}^{A}(t_{1},t')\Big).\label{eqm1}
\end{multline}
The Green's function, Hamiltonian $h$, and self-energies are written in the Kadanoff-Baym equations as matrices in molecular space and the molecular orbital indices are omitted here and  in the subsequent derivations for brevity. Here $h(t)$ means $h(x(t))$.

To transform these equations into the Wigner space, we define the central time $T$ and relative time $\tau$ as 
\begin{equation}
T=\frac{1}{2}(t + t'),
\end{equation}
and 
\begin{equation}
\tau=t-t',
\end{equation}
and introduce the Wigner transform of an arbitrary Green's function component 
\begin{equation}
\widetilde{\mathcal{G}}(T,\omega)=\int  d\tau e^{i\omega\tau}\mathcal{G}(t,t^{\prime}).
\end{equation}
The inverse Wigner transform from the Wigner space to the time domain takes the form 
\begin{equation}
\mathcal{G}(t,t')=\frac{1}{2\pi}\int  d\omega e^{-i\omega\tau}\widetilde{\mathcal{G}}(T,\omega).
\end{equation}
Applying the Wigner transform to both sides of (\ref{eqm2}) and (\ref{eqm1}) yields the Kadanoff-Baym equations of motion in the Wigner space 
\begin{multline}
\Big(\omega+\frac{i}{2}\partial_{T}-e^{\frac{1}{2i}\partial_{\omega}^{\mathcal{G}}d_{T}^{h}} h(t)\Big)\widetilde{\mathcal{G}}^{R/A}(T,\omega)=I\\
+e^{\frac{1}{2i}(\partial_{T}^{\Sigma}\partial_{\omega}^{\mathcal{G}}-\partial_{\omega}^{\Sigma}\partial_{T}^{\mathcal{G}})}\widetilde{\Sigma}^{R/A}(T,\omega)\widetilde{\mathcal{G}}^{R/A}(T,\omega),\label{eqm4}
\end{multline}
and 
\begin{multline}
\Big(\omega+\frac{i}{2}\partial_{T}-e^{\frac{1}{2i}\partial_{\omega}^{\mathcal{G}}d_{T}^{h}} h(t)\Big)\widetilde{\mathcal{G}}^{<}(T,\omega)=e^{\frac{1}{2i}(\partial_{T}^{\Sigma}\partial_{\omega}^{\mathcal{G}}-\partial_{\omega}^{\Sigma}\partial_{T}^{\mathcal{G}})}\\
\times\Big(\widetilde{\Sigma}^{R}(T)\widetilde{\mathcal{G}}^{<}(T,\omega)+\widetilde{\Sigma}^{<}(T,\omega)\widetilde{\mathcal{G}}^{A}(T,\omega)\Big).\label{eqm3}
\end{multline}
The equations of motion in the Wigner space are solved by treating the time derivatives with respect to the central time as a small parameter. This treatment means that we treat changes in the self-energies and Green's functions as slow with respect to central time and fast with respect to relative time. Central time dependence arises through the classical variable $x(t)$ and so the slow variation with respect to central time is associated with slow nuclear dynamics. The relative time is associated with the electronic time-scale and, in our case, the characteristic tunneling time for the electron to
transport across the molecule. Therefore the small parameter in our theory will be the ratio between the characteristic time-scales of nuclear motion and electron tunneling. The tunneling timescale can be estimated as $1/\Gamma$ where $\Gamma$ is the molecular level broadening due to the molecule-lead coupling. The time-scale for nuclear dynamics is given by $1/\Omega$ where $\Omega$ is the characteristic frequency for nuclear motion. Therefore the small parameter in our theory is $\frac{\Omega}{\Gamma}$.

The solution described below follows closely the ideas of previous authors\cite{Bode12,Dou2017,subotnik18,kershaw17,kershaw18,kershaw19}. The exponential operators in  Eqs.(\ref{eqm4},\ref{eqm3}) are expanded up to the first order in the time derivatives, where we result in a truncated equation of motion for the retarded, advanced and lesser components of the Green's functions
\begin{multline}
\Big(\omega+\frac{i}{2}\partial_{T}-\Big[1+\frac{1}{2i}\partial_{\omega}^{\mathcal{G}}d_{T}^{h}\Big]h\Big)\widetilde{\mathcal{G}}^{R/A}\\= 
\
I + \widetilde{\Sigma}^{R/A}\widetilde{\mathcal{G}}^{R/A}
\
+\frac{1}{2i}\Big(\partial_{T}^{\Sigma}\partial_{\omega}^{\mathcal{G}}-\partial_{\omega}^{\Sigma}\partial_{T}^{\mathcal{G}}\Big)\widetilde{\Sigma}^{R/A}\widetilde{\mathcal{G}}^{R/A},
\label{perturbedR/A}
\end{multline}
and
\begin{multline}
\Big(\omega+\frac{i}{2}\partial_{T}-\Big[1+\frac{1}{2i}\partial_{\omega}^{\mathcal{G}}d_{T}^{h}\Big]h\Big)\widetilde{\mathcal{G}}^{<}\\
\
=\widetilde{\Sigma}^{R}\widetilde{\mathcal{G}}^{<}+\widetilde{\Sigma}^{<}\widetilde{\mathcal{G}}^{A}
\
+\frac{1}{2i}\Big(\partial_{T}^{\Sigma}\partial_{\omega}^{\mathcal{G}}-\partial_{\omega}^{\Sigma}\partial_{T}^{\mathcal{G}}\Big)\Big(\widetilde{\Sigma}^{R}\widetilde{\mathcal{G}}^{<}+\widetilde{\Sigma}^{<}\widetilde{\mathcal{G}}^{A}\Big).
\label{perturbedL}
\end{multline}
Here  the function notation of the self-energies and Green's functions  have been suppressed  for brevity. We now solve each of the
equations above separately: first considering the retarded/advanced equation of motion followed by the lesser equation. Finding solutions with the derivatives up to the first order requires perturbative expansions of both the Green's functions and self-energies. In doing so, we expand all Green's function and self-energy components into a power series in terms of the small parameter:
\begin{equation}
\widetilde{\mathcal{G}}=\widetilde{\mathcal{G}}_{(0)} +\widetilde{\mathcal{G}}_{(1)} + O\left(\frac{\Omega}{\Gamma}\right)^2, \label{ansatzR/A}
\end{equation}
and
\begin{equation}
\widetilde{\Sigma}=\widetilde{\Sigma}_{(0)}+ \widetilde{\Sigma}_{(1)} + O\left(\frac{\Omega}{\Gamma}\right)^2.
\label{SelfEnergyExpansion/A}
\end{equation}

Here the terms $\widetilde{\mathcal{G}}_{(0)}$ and $\widetilde{\Sigma}_{(0)}$ depend on the instantaneous
nuclear geometry $x(t)$ only, while $\widetilde{\mathcal{G}}_{(1)}$ and $\widetilde{\Sigma}_{(1)}$ depend
on the nuclear geometry and are linear in velocities $\dot x(t)$.

Substituting (\ref{ansatzR/A}) and (\ref{SelfEnergyExpansion/A}) into (\ref{perturbedR/A}) and splitting the equation
based on order results in the equations 
\begin{equation}
\Big(\omega-h\Big)\widetilde{\mathcal{G}}_{(0)}^{R/A}=I + \widetilde{\Sigma}_{(0)}^{R/A}\widetilde{\mathcal{G}}_{(0)}^{R/A},
\end{equation}
and 
\begin{multline}
\Big(\omega-h\Big)\widetilde{\mathcal{G}}_{(1)}^{R/A}-\frac{1}{2i}\partial_{T} \widetilde{\mathcal{G}}_{(0)}^{R/A}-\frac{1}{2i}d_{T}h\partial_{\omega}\widetilde{\mathcal{G}}_{(0)}^{R/A}\\=
\widetilde{\Sigma}_{(0)}^{R/A}\widetilde{\mathcal{G}}_{(1)}^{R/A}
+ \widetilde{\Sigma}_{(1)}^{R/A}\widetilde{\mathcal{G}}_{(0)}^{R/A}
\\
+\frac{1}{2i}\Big(\partial_{T}^{\Sigma} \partial_{\omega}^{\mathcal{G}}-\partial_{\omega}^{\Sigma}\partial_{T}^{\mathcal{G}}\Big)\widetilde{\Sigma}_{(0)}^{R/A} \widetilde{\mathcal{G}}_{(0)}^{R/A}.
\label{Afirst}
\end{multline}
The equation for the zeroth order is easily solved to give 
\begin{equation}
\widetilde{\mathcal{G}}_{(0)}^{R/A}=\Big(\omega-h-\widetilde{\Sigma}_{(0)}^{R/A}\Big)^{-1}=G^{R/A},
\end{equation}
which is the standard, adiabatic retarded/advanced Green's function $G^{R/A}$. Considering now the first order equation of motion (\ref{Afirst}), we rearrange in terms of $\widetilde{\mathcal{G}}_{(1)}^{R/A}$ to obtain 
\begin{multline}
\widetilde{\mathcal{G}}_{(1)}^{R/A}= G^{R/A} \widetilde{\Sigma}_{(1)}^{R/A}  G^{R/A} \\
+ \frac{1}{2i}G^{R/A}\Big(\mathcal{A}^{R/A}\partial_{T}+ \mathcal{B}^{R/A} \partial_{\omega}\Big)G^{R/A}.
\label{GF_FirstOrder/A}
\end{multline}

Here we have defined the quantities $\mathcal{A}^{R/A}=I-\partial_{\omega}\widetilde{\Sigma}_{(0)}^{R/A}$ and $\mathcal{B}^{R/A}=\partial_{T}h + \partial_{T}\widetilde{\Sigma}_{(0)}^{R/A}$
in the interest of brevity, a convention that will be used for the
remainder of this derivation. 

The first order Green's function derivatives are found to be

\begin{equation}
\partial_{\omega} G^{R/A} = - G^{R/A} \mathcal{A}^{R/A} G^{R/A},
\end{equation}
and
\begin{equation}
\partial_{T} G^{R/A} = G^{R/A} \mathcal{B}^{R/A} G^{R/A}.
\end{equation}

This enables us to simplify (\ref{GF_FirstOrder/A}) to
\begin{multline}
\widetilde{\mathcal{G}}_{(1)}^{R/A} = G^{R/A} \widetilde{\Sigma}_{(1)}^{R/A}  G^{R/A} \\
\
+ \frac{1}{2i} G^{R/A} \Big[\mathcal{A}^{R/A} G^{R/A}, \mathcal{B}^{R/A}G^{R/A} \Big]_{-}.
\end{multline}
Notice that  that  the Green's functions, self-energies, and their derivatives will become scalars for transport in the single molecular energy level case and therefore the commutator term will vanish.

We now consider the equation of motion for the lesser Green's function which is given by (\ref{perturbedL}). The expansions (\ref{ansatzR/A}) and (\ref{SelfEnergyExpansion/A}) are substituted into (\ref{perturbedL}) and, as before, split based on order to give
\begin{equation}
\Big(\omega-h\Big)\widetilde{\mathcal{G}}_{(0)}^{<}=\widetilde{\Sigma}_{(0)}^{R}\widetilde{\mathcal{G}}_{(0)}^{<}+\widetilde{\Sigma}_{(0)}^{<}G^{A},
\end{equation}
and 
\begin{multline}
\Big(\omega-h\Big)\widetilde{\mathcal{G}}_{(1)}^{<}+\frac{i}{2}\partial_{T}\widetilde{\mathcal{G}}_{(0)}^{<}+\frac{i}{2}d_{T}h\partial_{\omega}\mathcal{G}^{<}\\
\
= \widetilde{\Sigma}_{(0)}^{R}\widetilde{\mathcal{G}}_{(1)}^{<} + \widetilde{\Sigma}_{(0)}^{<}\widetilde{\mathcal{G}}_{(1)}^{A}
\
+ \widetilde{\Sigma}_{(1)}^{R}\widetilde{\mathcal{G}}_{(0)}^{<} + \widetilde{\Sigma}_{(1)}^{<} G^{A}\\
\
+\frac{1}{2i}\Big(\partial_{T}^{\Sigma} \partial_{\omega}^{\mathcal{G}}-\partial_{\omega}^{\Sigma} \partial_{T}^{\mathcal{G}}\Big) \Big( \widetilde{\Sigma}_{(0)}^{R} \widetilde{\mathcal{G}}_{(0)}^{<} +  \widetilde{\Sigma}_{(0)}^{<} \widetilde{\mathcal{G}}_{(0)}^{A} \Big).
\label{GF_First_step/L}
\end{multline}

The zeroth order equation is easily solved to give 
\begin{equation}
\widetilde{\mathcal{G}}_{(0)}^{<}=G^{R}\widetilde{\Sigma}_{(0)}^{<}G^{A}=G^{<},
\end{equation}
which is again the standard expression for the adiabatically computed  lesser Green's function
$G^{<}$. Considering now the first order equation, we first compute explicit expressions for the lesser Green's function derivatives:

\begin{equation}
\partial_{\omega}G^{<}=-G^{R}\mathcal{A}^{R}G^{<}-G^{<}\mathcal{A}^{A}G^{A}+G^{R}\partial_{\omega}\widetilde{\Sigma}_{(0)}^{<}G^{A},
\end{equation}
and
\begin{equation}
\partial_{T}G^{<}=G^{R}\mathcal{B}^{R}G^{<}+G^{<}\mathcal{B}^{A}G^{A}+G^{R}\partial_{T}\widetilde{\Sigma}_{(0)}^{<}G^{A}.
\end{equation}

By substituting these expressions into (\ref{GF_First_step/L}) and rearranging the equation in terms of $\widetilde{\mathcal{G}}_{(1)}^{<}$, we find

\begin{widetext}

\begin{multline}
\widetilde{\mathcal{G}}_{(1)}^{<}
\
= G^{R}\widetilde{\Sigma}_{(0)}^{<}\widetilde{\mathcal{G}}_{(1)}^{A}
+ G^{R}\widetilde{\Sigma}_{(1)}^{<}G^{A}
+ G^{R}\widetilde{\Sigma}_{(1)}^{R}G^{<}
\
+ \frac{1}{2i} \Big[G^{R} \mathcal{A}^{R}, G^{R} \mathcal{B}^{R} \Big]_{-}G^{<} \\
\
\\
+ \frac{1}{2i}  G^{R}  \Big\{\mathcal{A}^R G^< \mathcal{B}^A
+ \mathcal{A}^R G^R \partial_{T} \widetilde{\Sigma}_{(0)}^<
+ \mathcal{B}^R G^R \partial_{\omega} \widetilde{\Sigma}_{(0)}^< + \text{h.c.} \Big\} G^A.
\end{multline}

\end{widetext}

It remains to calculate the expressions for our self-energies. Beginning  with our general definition for the self-energy in real-time, we then apply the Wigner transform which allows us to extract an adiabatic case along with our first non-adiabatic correction.
The exact self-energy computed in real time is\cite{haug-jauho}
\begin{equation}
\Sigma(ct,c't')=\sum_{k\alpha}v_{ck\alpha}(t)\mathcal{G}_{0,k\alpha}(t,t')v_{k\alpha c'}(t') \label{SigmaRealTime},
\end{equation}
where $ \mathcal{G}_{0,k\alpha}(t,t')$ is the non-interacting Green's functions for the separated leads.
Application of the Wigner transform to (\ref{SigmaRealTime}) yields
\begin{equation}
\widetilde{\Sigma}(c,c') = \sum_{k\alpha} e^{\frac{1}{2i} (\partial_{T}^{v} - \partial_{T}^{v'})\partial_{\omega}^{\mathcal{G}}}v_{ck\alpha}(T) \widetilde{\mathcal{G}}_{0,k\alpha}(T,\omega)v_{k\alpha c'}(T),
\end{equation}
where we have used $\partial_{T}^{v}$ and $\partial_{T}^{v'}$ to denote the central-time derivative with respect to $v_{k\alpha c}$ and $v_{k\alpha c'}$ respectively. Expansion of the above exponential enables us to partition the equation into orders of magnitude of $\partial_{T}$ such that we find our adiabatic self-energy and corrections as 

\begin{equation}
\widetilde{\Sigma}_{(0)}(c,c') = \sum_{k\alpha} v_{ck\alpha}(T) 
\widetilde{\mathcal{G}}_{0,k\alpha}(T,\omega)v_{k\alpha c'}(T),
\end{equation}
and
\begin{equation}
\widetilde{\Sigma}_{(1)}(c,c') = \frac{\dot x}{2i}\partial_{\omega}\Big(\widetilde{\Psi}_{(0)}(T,\omega) - \widetilde{\Phi}_{(0)}(T,\omega)\Big). 
\end{equation}

Here we have introduced the self-energy-like quantities:
\begin{equation}
\label{psi}
\Psi_{cc'}(t,t') = \sum_{\substack{k \alpha}} \Lambda_{c k \alpha}(t) \mathcal{G}_{0,k \alpha}(t,t') v_{k \alpha c'}(t'),
\end{equation}
and
\begin{equation}
\label{phi}
\Phi_{cc'}(t,t')  = \sum_{\substack{k \alpha}} v_{c k \alpha}(t) \mathcal{G}_{0,k \alpha}(t,t') \Lambda_{k \alpha c'}(t'),
\end{equation}
where $\widetilde{\Psi}$ and $\widetilde{\Phi}$ are their respective Wigner transformations. The  derivative of the tunneling amplitude with respect to our classical coordinate is defined as
\begin{equation}
\Lambda_{k \alpha c} = \frac{dv_{k \alpha c}}{dx}.
\label{lambda}
\end{equation}

By enacting an equivalent derivation as for the self-energy, we can decompose these new quantities into an adiabatic term and a first-order correction as given by:

\begin{equation}
\widetilde{\Psi}_{(0)}(c,c') = \sum_{k\alpha} \Lambda_{ck\alpha}(T) 
\widetilde{\mathcal{G}}_{0,k\alpha}(T,\omega)v_{k\alpha c'}(T),
\end{equation}

\begin{equation}
\widetilde{\Phi}_{(0)}(c,c') = \sum_{k\alpha} v_{ck\alpha}(T) 
\widetilde{\mathcal{G}}_{0,k\alpha}(T,\omega)\Lambda_{k\alpha c'}(T),
\end{equation}

\begin{equation}
\widetilde{\Psi}_{(1)}(c,c') = \frac{1}{2i}\Big(
\
\partial^2_{T\omega} \widetilde{\Psi}_{(0)}(c,c') - 2 \partial_{\omega} \widetilde{\Omega}_{(0)}(c,c') \Big),
\end{equation}
and

\begin{equation}
\widetilde{\Phi}_{(1)}(c,c') = -\frac{1}{2i}\Big(
\
\partial^2_{T\omega} \widetilde{\Phi}_{(0)}(c,c') - 2 \partial_{\omega} \widetilde{\Omega}_{(0)}(c,c') \Big),
\end{equation}

where

\begin{equation}
\widetilde{\Omega}_{(0)}(c,c') = \sum_{k\alpha}\Lambda_{ck\alpha}(T)\widetilde{\mathcal{G}}_{k\alpha}(T,\omega)\Lambda_{k\alpha c'}(T).
\end{equation}

\subsection{Dynamical Corrections to Time-Dependent Electric Current}
In section II-B, we obtained first-order dynamical corrections to the retarded, advanced, and lesser central region Green's functions.  Let us now obtain an expression for the current that includes  first order dynamical  corrections due to the motion of interfacial atoms.
We begin with the general expression for the electric current flowing into the molecule from the $\alpha$ lead at time $t$ \cite{haug-jauho}:

\begin{multline}
I_{\alpha}(t)=\int  dt_{1}\text{Tr}\Big\{\mathcal{G}^{<}(t,t_{1})\Sigma_{\alpha}^{A}(t_{1},t)
+\mathcal{G}^{R}(t,t_{1})\Sigma_{\alpha}^{<}(t_{1},t)\\
- \Sigma^{A}_{\alpha}(t,t_1) \mathcal{G}^{<}(t_1,t)
- \Sigma^{<}_{\alpha}(t,t_1) \mathcal{G}^{R}(t_1,t)
\Big\}.
\end{multline}

To facilitate a transformation to the Wigner space, we introduce the two-time function $\mathcal{I}_{\alpha}(t,t')$
\begin{multline}
\mathcal{I}_{\alpha}(t,t')=\int  dt_{1}\text{Tr}\Big\{\mathcal{G}^{<}(t,t_{1})\Sigma_{\alpha}^{A}(t_{1},t')
+\mathcal{G}^{R}(t,t_{1})\Sigma_{\alpha}^{<}(t_{1},t')\\
- \Sigma^{A}_{\alpha}(t,t_1) \mathcal{G}^{<}(t_1,t')
- \Sigma^{<}_{\alpha}(t,t_1) \mathcal{G}^{R}(t_1,t')
\Big\},
\end{multline}
which becomes the electric current if $t=t'$
\begin{equation}
\mathcal{I}_{\alpha}(t,t) = I_{\alpha}(t).
\end{equation}
A transformation into the Wigner space yields

\begin{multline}
\widetilde{\mathcal{I}}_{\alpha} = \text{Tr} \Big\{
\
e^{\frac{1}{2i}(\partial_{T}^{\mathcal{G}}\partial_{\omega}^{\Sigma} - \partial_{\omega}^{\mathcal{G}}\partial_{T}^{\Sigma})}
\
\Big(\widetilde{\mathcal{G}}^{<} \widetilde{\Sigma}^{A}_{\alpha} + \widetilde{\mathcal{G}}^{R} \widetilde{\Sigma}^{<}_{\alpha}\Big)
\\\
- e^{\frac{1}{2i}(\partial_{T}^{\Sigma}\partial_{\omega}^{\mathcal{G}} - \partial_{\omega}^{\Sigma}\partial_{T}^{\mathcal{G}})}
\
\Big(\widetilde{\Sigma}^{<}_{\alpha}\widetilde{\mathcal{G}}^{A} +  \widetilde{\Sigma}^{R}_{\alpha} \widetilde{\mathcal{G}}^{<}\Big)
\Big\}.
\
\end{multline}

Following the ideas of the previous section, we now expand the exponential operators up to the first order, along with including our non-adiabatic corrections to the Green's functions and self-energies. We then perform an inverse Wigner transform back to real time, in which we set $t=t'$. This yields the equation for our current in real time in terms of an adiabatic component and a first order correction. The adiabatic component is found to be 

\begin{multline}
I_{\alpha}^{(0)}=\frac{1}{2\pi} \int  d\omega
\
\text{Tr}\Big\{
\
G^{<}\widetilde{\Sigma}_{\alpha (0)}^{A} + G^{R}\widetilde{\Sigma}_{\alpha (0)}^{<} 
\\\
- \widetilde{\Sigma}_{\alpha (0)}^{<} G^{A} - \widetilde{\Sigma}_{\alpha (0)}^{R} G^{<} 
\Big\},
\label{adiabaticCurrent}
\end{multline}

while the first order correction is given by

\begin{multline}
I_{\alpha}^{(1)}=\frac{1}{2\pi} \int  d\omega
\
\text{Tr}\Big\{
\
\widetilde{\mathcal{G}}_{(1)}^{<} \widetilde{\Sigma}_{\alpha (0)}^{A}
\
+ \widetilde{\mathcal{G}}_{(1)}^{R} \widetilde{\Sigma}_{\alpha (0)}^{<}
\\\
- \widetilde{\Sigma}_{\alpha (0)}^{<} \widetilde{\mathcal{G}}_{(1)}^{A}
\
- \widetilde{\Sigma}_{\alpha (0)}^{R} \widetilde{\mathcal{G}}_{(1)}^{<}
\
+ \widetilde{\mathcal{G}}_{(0)}^{<} \widetilde{\Sigma}_{\alpha (1)}^{A}
\\\
+ \widetilde{\mathcal{G}}_{(0)}^{R} \widetilde{\Sigma}_{\alpha (1)}^{<}
\
- \widetilde{\Sigma}_{\alpha (1)}^{<} \widetilde{\mathcal{G}}_{(0)}^{A}
\
- \widetilde{\Sigma}_{\alpha (1)}^{R} \widetilde{\mathcal{G}}_{(0)}^{<}
\
\\\
+ \frac{1}{2i} \Big( \partial_T G^< \partial_{\omega} \widetilde{\Sigma}_{\alpha (0)}^{A}
- \partial_{\omega}G^{<} \partial_T \widetilde{\Sigma}_{\alpha (0)}^{A} 
+ \partial_T G^R \partial_{\omega} \widetilde{\Sigma}_{\alpha (0)}^{<} \\
- \partial_{\omega}G^{R} \partial_T \widetilde{\Sigma}_{\alpha (0)}^{<} 
+ \partial_T \widetilde{\Sigma}_{\alpha (0)}^< \partial_{\omega} G^A
- \partial_{\omega} \widetilde{\Sigma}_{\alpha (0)}^< \partial_T G^A \\
+ \partial_T \widetilde{\Sigma}_{\alpha (0)}^R \partial_{\omega} G^<
- \partial_{\omega} \widetilde{\Sigma}_{\alpha (0)}^R \partial_T G^<
\Big) \Big\}.
\label{firstOrderCurrent}
\end{multline}

We can simplify (\ref{adiabaticCurrent}) and (\ref{firstOrderCurrent}) by utilising the following identities:

\begin{equation}
\Big(\widetilde{\mathcal{G}}_{(0,1)}^R\Big)^\dag = \widetilde{\mathcal{G}}_{(0,1)}^A, \;\;\;\;
\Big(G^<\Big)^\dag = - G^<,
\label{identity12}
\end{equation} 
and
\begin{equation}
\Big(\widetilde{\Sigma}_{(0,1)}^R\Big)^\dag = \widetilde{\Sigma}_{(0,1)}^A, \;\;\;\;
\Big(\widetilde{\Sigma}_{(0,1)}^<\Big)^\dag = - \widetilde{\Sigma}_{(0,1)}^<.
\end{equation}

Our final general expressions for the current are then given by

\begin{multline}
I_{\alpha}^{(0)}=\frac{1}{\pi} \int  d\omega
\
\text{ReTr}\Big\{
\
G^{<}\widetilde{\Sigma}_{\alpha (0)}^{A} + G^{R}\widetilde{\Sigma}_{\alpha (0)}^{<} \Big\}, 
\end{multline}

and

\begin{multline}
I_{\alpha}^{(1)}=\frac{1}{2\pi} \int  d\omega
\
\text{Tr}\Big\{
\
\widetilde{\mathcal{G}}_{(1)}^{<} \widetilde{\Sigma}_{\alpha (0)}^{A}
\
- \widetilde{\Sigma}_{\alpha (0)}^{R} \widetilde{\mathcal{G}}_{(1)}^{<} \Big\}
\
\\\
+ \frac{1}{\pi} \int  d\omega
\
\text{ReTr}\Big\{\widetilde{\mathcal{G}}_{(1)}^{R} \widetilde{\Sigma}_{\alpha (0)}^{<}
\
+ \widetilde{\mathcal{G}}_{(0)}^{<} \widetilde{\Sigma}_{\alpha  (1)}^{A}
\
+ \widetilde{\mathcal{G}}_{(0)}^{R} \widetilde{\Sigma}_{\alpha  (1)}^{<}
\\\
+ \frac{1}{2i} \Big( \partial_T G^< \partial_{\omega} \widetilde{\Sigma}_{\alpha (0)}^{A}
- \partial_{\omega} G^{<} \partial_T \widetilde{\Sigma}_{\alpha (0)}^{A} 
+ \partial_T G^R \partial_{\omega} \widetilde{\Sigma}_{\alpha (0)}^{<} \\
- \partial_{\omega}G^{R} \partial_T \widetilde{\Sigma}_{\alpha (0)}^{<} 
\Big) \Big\}.
\end{multline}

\subsection{Current-induced forces: adiabatic, viscous, and random component}

We have obtained analytical solutions to the Kadanoff-Baym equations for Green's functions as functions of instantaneous positions and velocities. Therefore, for any given trajectory we know how to compute all system observables.
Our aim here is to derive a Langevin-like equation  to obtain the stochastic trajectory for the molecular junction geometry  $x=x(t)$. Our derivation follows the ideas introduced by von Oppen et al.\cite{Bode12}, then later expanded upon by Subotnik and Dou \cite{Dou2017,subotnik18}.

The derivation starts on a purely quantum-mechanical footing, by considering quantum position and momentum  operators ($\hat x$ and 
$\hat p$) which correspond to the classical variable $x$.
The Heisenberg equation of evolution for the momentum operator gives the expression for the quantum force
\begin{equation}
\hat{f}(t)  = i \Big[\hat{H}(t), \hat{p} \Big]_{-},
\end{equation}
where $\hat{H}(t)$ is the full Hamiltonian of the system. 
 Note that in contrast to previous sections, we have been careful to make explicit the operator notation so that the quantum and classical quantities are easily distinguishable. In the coordinate representation,  $\hat{p} = - i \hbar \partial_x$; therefore
\begin{equation}
\hat{f}(t) = - \partial_x \hat{H}(t),
\end{equation}
which when making a substitution for $\hat{H}(t)$  (\ref{hamiltonian}) becomes
\begin{equation}
\hat{f}(t) = - U^{\prime} - \partial_x \hat{H}_M(t) - \partial_x \hat{H}_{LM}(t) - \partial_x \hat{H}_{RM}(t),
\label{yeehat}
\end{equation}
where we have added the classical potential $U$ to the Hamiltonian where $U^{\prime} = \partial_{x} U$. Making a substitution for each Hamiltonian results in 
\begin{equation}
\hat{f}(t) =  -  U^{\prime}
- \sum_{ij} \partial_x h_{ij} \hat{a}_i^{\dag} \hat{a}_j
 - \sum_{\substack{k\alpha i}} \Big( \Lambda_{i k \alpha} \hat{a}^{\dagger}_{i} \hat{a}_{k \alpha} +  \Lambda_{k \alpha i} \hat{a}^{\dagger}_{k \alpha} \hat{a}_{i} \Big),
\label{yee}
\end{equation} 
where the quantity $\Lambda$ is given by (\ref{lambda}). First, we compute the average of the force operator at time $t$ and then we introduce the stochastic fluctuations around this average. 
The force exerted by electrons is
\begin{equation}
\hat F = \hat f +U'
\end{equation}
and its average can be conveniently expressed in terms of Green's functions
\begin{multline}
F(t) = 
i \sum_{ij} \partial_x h_{ij}(t) \mathcal{G}_{ji}^{<}(t,t)
\\
+ i \sum_{\substack{k\alpha i}} \Big( \Lambda_{i k \alpha}(t) \mathcal{G}_{k \alpha i}^{<}(t,t) 
\
+ \Lambda_{k \alpha i}(t) \mathcal{G}_{ik \alpha}^{<}(t,t)  \Big).
\end{multline} 
 We now introduce our auxiliary two-time function $\mathcal{F}(t,t')$ as
\begin{multline}
\mathcal{F}(t,t') = 
i \sum_{ij} \partial_x h_{ij}(t) \mathcal{G}_{ji}^{<}(t,t')
\\
+ i \sum_{\substack{k\alpha i}} \Big( \Lambda_{i k \alpha}(t) \mathcal{G}_{k \alpha i}^{<}(t,t') 
\
+ \Lambda_{k \alpha i}(t) \mathcal{G}_{ik \alpha}^{<}(t,t')  \Big),
\end{multline} 
which has the property $\mathcal{F}(t,t) = F(t)$. As previously, we first utilize the Dyson equation for our Green's functions spanning the molecular space and the leads . In doing so, we express $\mathcal{F}(t,t') $ in terms of molecular space quantities:
\begin{multline}
\mathcal{F}(t,t') = i \sum_{ij} \partial_x h_{ij} \mathcal{G}_{ji}^{<}(t,t')
\\\
+ i \sum_{ij} \int   dt_1 \Big(\mathcal{G}_{ij}^{<}(t,t_1) \Phi_{\alpha ,ji}^A (t_1,t')
\
+\mathcal{G}_{ij}^{R}(t,t_1) \Phi_{\alpha ,ji}^< (t_1,t')
\\\
+ \Psi_{\alpha ,ij}^< (t,t_1) \mathcal{G}_{ji}^A(t_1,t')
\
+ \Psi_{\alpha ,ij}^R (t,t_1) \mathcal{G}_{ji}^<(t_1,t')
\Big),
\end{multline}
where  our self-energy-like quantities (\ref{phi}) and (\ref{psi}) have appeared once again in the equation. Taking a trace over the molecular states and performing a transformation into the Wigner space yields
\begin{multline}
\widetilde{\mathcal{F}} = \text{Tr}\Big\{ie^{\frac{1}{2i} \partial_T^h \partial_{\omega}^{\mathcal{G}}} \partial_x h \widetilde{\mathcal{G}}^<
\\\
+ ie^{\frac{1}{2i}(\partial_T^{\Psi} \partial_{\omega}^{\mathcal{G}} - \partial_{\omega}^{\Psi} \partial_T^{\mathcal{G}})}
\
\Big(\widetilde{\Psi}^< \widetilde{\mathcal{G}}^A + \widetilde{\Psi}^R \widetilde{\mathcal{G}}^< \Big)
\\\
+ ie^{\frac{1}{2i}(\partial_T^\mathcal{G} \partial_{\omega}^{\Phi} - \partial_{\omega}^{\mathcal{G}} \partial_T^{\Phi})}
\
\Big(\widetilde{\mathcal{G}}^< \widetilde{\Phi}^A + \mathcal{G}^R \widetilde{\Phi}^< \Big)\Big\}.
\end{multline}

Now, by taking the inverse Wigner transform and letting $t = t'$ such that $\mathcal{F}(t,t) = F(t)$, we decompose our classical force into an adiabatic component and a velocity-dependent correction (which will correspond to our viscosity force). The adiabatic component of the force is given by
\begin{multline}
F_{(0)}(t) = \text{Tr}\Big\{
\
\frac{1}{2\pi} \int   d\omega i \partial_x h G^<
\\\
- \frac{1}{\pi} \int   d\omega\text{Im}
\
\Big(\widetilde{\Psi}_{(0)}^< G^A + \widetilde{\Psi}_{(0)}^R G^<
\Big)\Big\},
\label{xx}
\end{multline}
while the velocity dependent first order component is given by
\begin{multline}
F_{(1)}(t) = \text{Tr}\Big\{
\
\frac{1}{2\pi} \int   d\omega \Big( i\partial_x h \widetilde{\mathcal{G}}_{(1)}^< + \frac{v}{2} d^2_x h \partial_{\omega} G^< \Big)
\\
+\frac{i}{2\pi} \int   d\omega \Big(
\
\widetilde{\Psi}_{(1)}^< G^A + \widetilde{\Psi}_{(1)}^R G^<
+ \widetilde{\Psi}_{(0)}^< \widetilde{\mathcal{G}}_{(1)}^A + \widetilde{\Psi}_{(0)}^R \widetilde{\mathcal{G}}_{(1)}^<
\\\
+ G^< \widetilde{\Phi}_{(1)}^A + G^R \widetilde{\Phi}_{(1)}^<
+ \widetilde{\mathcal{G}}_{(1)}^< \widetilde{\Phi}_{(0)}^A + \widetilde{\mathcal{G}}_{(1)}^R \widetilde{\Phi}_{(0)}^< \Big)
\\\
+\frac{1}{4\pi} \int   d\omega \Big(
\
\partial_T \widetilde{\Psi}_{(0)}^< \partial_{\omega} G^A
+ \partial_T \widetilde{\Psi}_{(0)}^R \partial_{\omega} G^<
- \partial_{\omega} \widetilde{\Psi}_{(0)}^< \partial_T G^A \\
- \partial_{\omega} \widetilde{\Psi}_{(0)}^R \partial_T G^<
\
+ \partial_T G^< \partial_{\omega} \widetilde{\Phi}_{(0)}^A
+ \partial_T G^R \partial_{\omega} \widetilde{\Phi}_{(0)}^< \\
- \partial_{\omega} G^< \partial_T \widetilde{\Phi}_{(0)}^A
- \partial_{\omega} G^R \partial_T \widetilde{\Phi}_{(0)}^<
\Big)\Big\}.
\end{multline}

We note that the second term reduces to zero as

\begin{equation}
\int_{-\infty}^{\infty}d\omega \partial_{\omega} G^< 
= 0.
\end{equation}

This can be further simplified through the use of (\ref{identity12})  along with the following identity:
\begin{equation}
\Big(\widetilde{\Phi}_{(0,1)}^R \Big)^{\dagger} = \widetilde{\Psi}_{(0,1)}^A.  
\end{equation}
Our final expression for the velocity-dependent force is then given by
\begin{multline}
F_{(1)}(t) = \text{Tr}\Big\{
\
\frac{1}{2\pi} \int   d\omega  i \partial_x h \widetilde{\mathcal{G}}_{(1)}^< 
\\
+\frac{i}{2\pi} \int   d\omega \Big(
\
\widetilde{\Psi}_{(0)}^R \widetilde{\mathcal{G}}_{(1)}^<
+ \widetilde{\mathcal{G}}_{(1)}^< \widetilde{\Phi}_{(0)}^A \Big)
\\\
-\frac{1}{\pi} \int   d\omega \text{Im} \Big\{
\
\widetilde{\Psi}_{(1)}^< G^A + \widetilde{\Psi}_{(1)}^R G^<
+ \widetilde{\Psi}_{(0)}^< \widetilde{\mathcal{G}}_{(1)}^A \Big\}
\\\
+ \frac{1}{2\pi}\int   d\omega\text{Re}\Big\{
\
\partial_T \widetilde{\Psi}_{(0)}^< \partial_{\omega} G^A
+ \partial_T \widetilde{\Psi}_{(0)}^R \partial_{\omega} G^< \\
- \partial_{\omega} \widetilde{\Psi}_{(0)}^< \partial_T G^A 
- \partial_{\omega} \widetilde{\Psi}_{(0)}^R \partial_T G^<
\Big\}\Big\}.
\label{xy}
\end{multline}
This expression is linear in velocity $\dot x$ and results in the viscosity force in the Langevin equation for $x(t)$.

To complete the Langevin equation, we need to define the diffusion coefficient as a time correlation of the force variations 
\begin{equation}
\langle \delta\hat  F(t) \delta \hat F(t')\rangle =D \delta(t-t'),
\end{equation}

where

\begin{equation}
 \delta\hat  F(t) = \hat  F(t) - \langle  \hat  F(t) \rangle.
\end{equation}

Through a tedious derivation, one can compute an explicit expression for the diffusion coefficient, a derivation that has been relegated to Appendix \ref{RWN}.  One can show that 
\begin{multline}
D(x) = \frac{1}{2\pi} \int   d\omega \text{Tr} \Big\{
\
\partial_x h G^> \partial_x h G^< 
\
+ G^> \widetilde{\Omega}^< 
+ \widetilde{\Omega}^> G^< 
\\\
+ 2 \text{Re} \Big[ \Big(\partial_x h + \widetilde{\Psi}^R + \widetilde{\Phi}^A\Big)
\
\Big(G^< \widetilde{\Psi}^> G^A + G^> \widetilde{\Psi}^< G^A + G^> \widetilde{\Psi}^R G^< \Big)
\\\
+ \widetilde{\Psi}^> G^A \widetilde{\Psi}^< G^A + \partial_x h G^< \widetilde{\Psi}^R G^> \Big] \Big\}. 
\label{D}
\end{multline}

The adiabatic force (\ref{xx}),  viscous force (\ref{xy}) and diffusion coefficient  (\ref{D})  are the main results of this section and will be used for modeling in subsequent sections.

\section{results}

\subsection{Model}
\label{model}

The molecular bridge is modeled by a single molecular orbital with
energy $\epsilon (x)$ as 
\begin{equation}
H_{M}=\epsilon (x) d^{\dag}d,
\end{equation}
where $x$ is a classical  time-dependent coordinate. In our case $x$ models a bond-length between the molecule and the left lead. This  $x$-dependence  of the molecular orbital comes from the voltage drop across the junction 
\begin{equation}
\epsilon (x) = \epsilon_0 + {\cal E} (x-x_0) + V_0,
\end{equation}
where 
\begin{equation}
{\cal E} = (\mu _L - \mu _R)/(L_L - L_R), 
\end{equation}
is the electric field across the junction
and
\begin{equation}
V_0 = \mu _L - L_L (\mu _L - \mu _R)/(L_L - L_R).  
\end{equation}
is the $x$-independent energy level shift.
Here we use $L_L$ and $L_R$ to denote the positions of the left and right leads, while $\mu _L$ and $\mu _R$ are the left and right lead chemical potentials. The equilibrium bond-length is denoted by $x_0$. The applied voltage bias $V$
will be   applied symmetrically $\mu_L=V/2$ and $\mu_R=-V/2$ in all our calculations.

We assume that the coupling to the right electrode is rigid and the coupling to the left depends on the bond-length:
\begin{eqnarray}
v_{k\alpha}(x)=\left\{ \begin{array}{c}
v_{L}s(x),\text{ if }\alpha=L\\
v_{R},\text{ if }\alpha=R
\end{array}\right.\label{vofx}
\end{eqnarray}
where the function $s(x)$ is taken in the form of the overlap between two 1s orbitals separated
by distance $ x$ as given by 
\begin{equation}
s( x)=e^{- x}(1+ x+ x^{2}/3),
\end{equation}
and $v_{L}$ and $v_{R}$ are two constants. This choice of coordinate
dependence mimics the behavior of a generic isotropic chemical bond.\cite{mcquarrie-qc}

We assume that the coupling to the left lead is time-dependent and the time-dependence comes from the variations of the bond-length between the molecule and the left lead. The choice of the left electrode is completely arbitrary, we can chose the linkage to the right electrode to be time-dependent as well.

We will use the wide-band approximation for the leads and,  in this limit, the leads self-energy components become 
\begin{equation}
\widetilde{\Sigma}_{(0)L}^{A}(T)=\frac{i}{2}\Gamma_{L}s^{2}(T),\;\;\;\;\;\widetilde{\Sigma}_{(0)R}^{A}=\frac{i}{2}\Gamma_{R},\label{sigmaAW}
\end{equation}
\begin{equation}
\widetilde{\Sigma}_{(0)L}^{R}(T)=-\frac{i}{2}\Gamma_{L}s^{2}(T),\;\;\;\;\;\widetilde{\Sigma}_{(0)R}^{R}(T)=-\frac{i}{2}\Gamma_{R},\label{sigmaRW}
\end{equation}
and 
\begin{equation}
\widetilde{\Sigma}_{(0)L}^{<}(T,\omega)=if_{L}(\omega)\Gamma_{L}s^{2}(T),\;\;\;\;\;\widetilde{\Sigma}_{(0)R}^{<}(T,\omega)=if_{R}(\omega)\Gamma_{R}.\label{sigma<W}
\end{equation}
Here $s(T)=s(x(T))$ and we have introduced the standard level broadening
function 
\begin{equation}
\Gamma_{\alpha}=2\pi|v_{\alpha}|^{2}\rho_{\alpha},
\end{equation}
where $\rho_{\alpha}$ is the density of single-particle states in lead $\alpha$.
Notice that the retarded/advanced self-energies for the left lead
have lost their energy dependence on $\omega$ in the wide-band limit
and retarded/advanced self-energies for the right lead become constants.

In the considered case of electron transport through a single resonant level,
the expressions for the non-adiabatic corrections can be further simplified since the 
Green's functions and self-energies are no longer matrices, in addition to the wide-band approximation killing some derivatives. 
The first order correction to the lesser Green's function becomes
\begin{multline}
\widetilde{\mathcal{G}}_{(1)}^{<}=- i G^{R} \text{Re} \Big\{G^< \mathcal{B}^A
+ G^R \partial_{T} \widetilde{\Sigma}_{(0)}^<
+ \mathcal{B}^R G^R \partial_{\omega} \widetilde{\Sigma}_{(0)}^< \Big\} G^A.
\end{multline}
It is expressed in terms of standard adiabatic (instantaneously computed along the nuclear trajectory $x(t)$) Green's functions
\begin{equation}
G^{A/R}=\left(\omega-\epsilon-\widetilde{\Sigma}^{A/R}\right)^{-1}, \;\;
G^{<}=G^{R}\widetilde{\Sigma}^{<} G^{A}. 
\end{equation}
The adiabatic electric current is
\begin{equation}
I_{\alpha}^{(0)}(t)=\frac{1}{\pi}\int d\omega \text{Re} \Big\{ G^{<}\widetilde{\Sigma}_{\alpha (0)}^{A}+G^{R}\widetilde{\Sigma}_{\alpha (0)}^{<} \Big\},
\end{equation}
while the first order velocity-dependent non-adiabatic correction to the electric current is
\begin{multline}
I_{\alpha}^{(1)}(t) = \frac{1}{\pi} \int   d\omega 
\
\text{Re} \Big\{\widetilde{\mathcal{G}}_{(1)}^{<} \widetilde{\Sigma}_{(0)}^A \Big\}
\\\
+ \frac{1}{2\pi} \int   d\omega \text{Im} \Big\{
\
\partial_T G^R \partial_{\omega} \widetilde{\Sigma}_{(0)}^< - \partial_{\omega} G^R \partial_T \widetilde{\Sigma}^< \Big\}.
\end{multline}

The adiabatic force is
\begin{multline}
F_{(0)} (t) = \frac{i {\cal E}}{2\pi} \int  d\omega  G^<
\\\
- \frac{1}{\pi} \int  d\omega \text{Im} \Big\{ \widetilde{\Psi}_{(0)}^< G^A + \widetilde{\Psi}_{(0)}^R G^< \Big\},
\label{f0model}
\end{multline}
and the dissipative force is given by
\begin{widetext}
\begin{multline}
F_{(1)} (t)  = \frac{i  {\cal E}}{2\pi} \int  d\omega  \mathcal{\widetilde{G}}_{(1)}^< 
\
- \frac{1}{\pi} \int  d\omega \text{Im} 
\
\Big\{ \widetilde{\Psi}_{(0)}^R \mathcal{\widetilde{G}}_{(1)}^< \Big\}
\
+\frac{1}{2\pi} \int  d\omega \text{Re} \Big\{
\
\partial_T \widetilde{\Psi}_{(0)}^< \partial_{\omega} G^A 
 - \partial{\omega} \widetilde{\Psi}_{(0)}^< \partial_T G^A \Big\}.
 \label{f1model}
\end{multline} 
Finally, the diffusion coefficient is
\begin{multline}
D(x) = \frac{1}{2\pi} \int   d\omega \Big(
\
{\cal E}^2 G^> G^< 
\
+ G^> \widetilde{\Omega}^< 
+ \widetilde{\Omega}^> G^< 
\\\
+ 2 \text{Re} \Big[ {\cal E}
\
\Big(G^< \widetilde{\Psi}^> G^A + G^> \widetilde{\Psi}^< G^A + 2G^> \widetilde{\Psi}^R G^< \Big)
\
+ \widetilde{\Psi}^> G^A \widetilde{\Psi}^< G^A \Big]\Big). 
\label{Dmodel}
\end{multline}
\end{widetext}

The  time-evolution of the bond-length $x$  is given
by the Langevin equation
\begin{equation}
m \ddot x = -U' + F_{(0)} (t) + \zeta(x) \dot x + \delta F(t).
\label{langevin}
\end{equation}
Here, the adiabatic force $F_{(0)} (t)$ is given by Eq.(\ref{f0model}),
the electronic viscosity $\zeta(x)$ is defined from velocity dependent contribution to the force (\ref{f1model}) as
\begin{equation}
 \xi(x) = -\frac{F_{(1)}(x)}{\dot x}, 
\end{equation}
and 
 $\delta F(t)$  is a white noise random force with diffusion coefficient  (\ref{Dmodel}). The classical potential $U(x)$ is taken to be harmonic 
\begin{equation}
U(x) = \frac{1}{2} k (x-x_0)^2,
\end{equation}
where $x_0$ is the equilibrium bond-length and $k$ is the spring constant associated with the bond strength.

\subsection{Calculations}

\begin{figure*}

\begin{subfigure}{0.32\textwidth}
\centering
\includegraphics[width=1\textwidth]{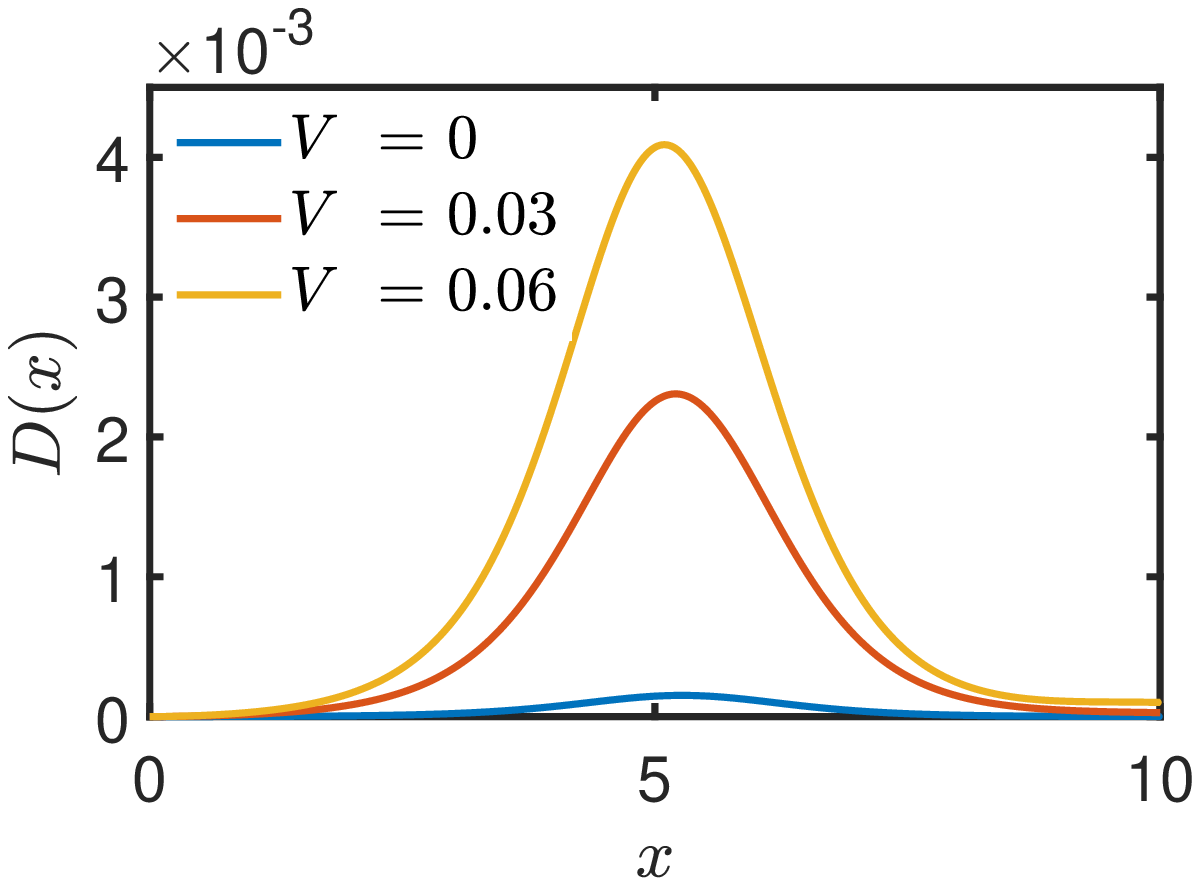}
\caption{}
\label{1a}
\end{subfigure}
\begin{subfigure}{0.32\textwidth}
\centering
\includegraphics[width=1\textwidth]{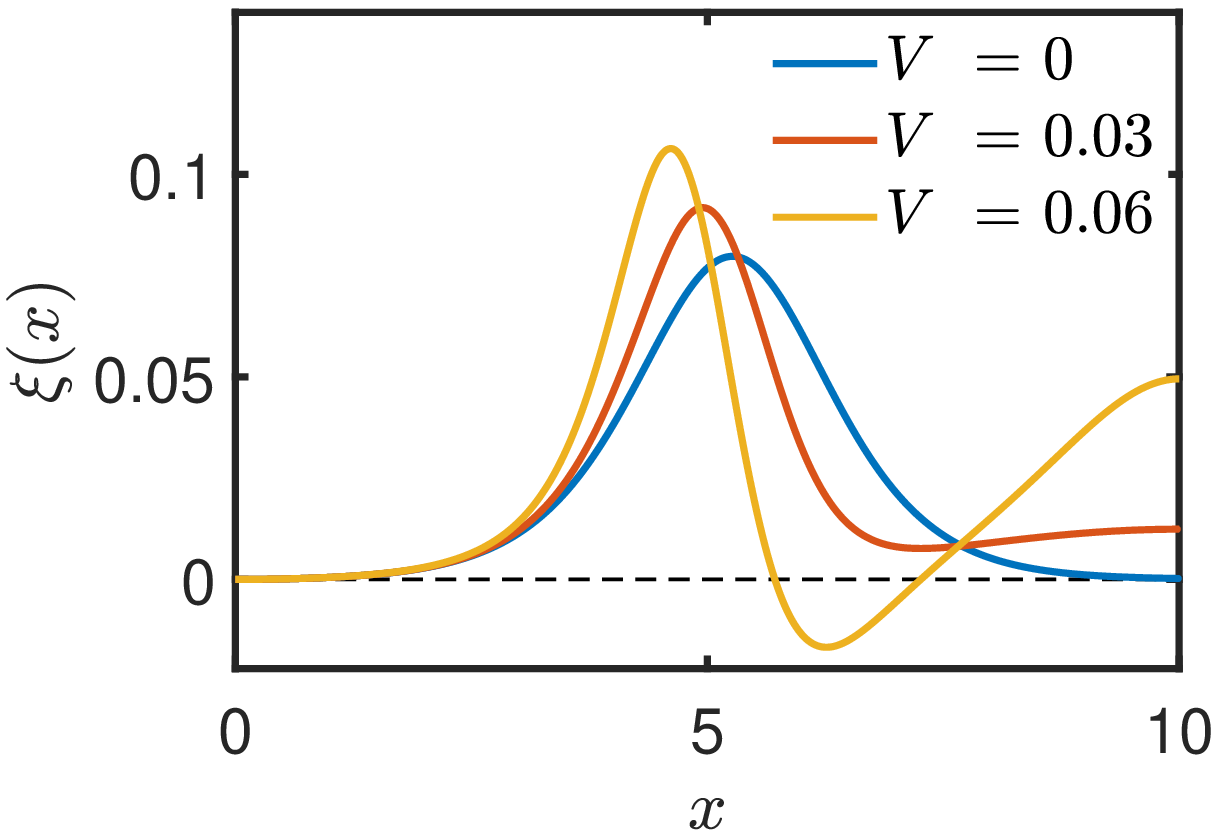}
\caption{}
\label{1b}
\end{subfigure}
\begin{subfigure}{0.32\textwidth}
\centering
\includegraphics[width=1\textwidth]{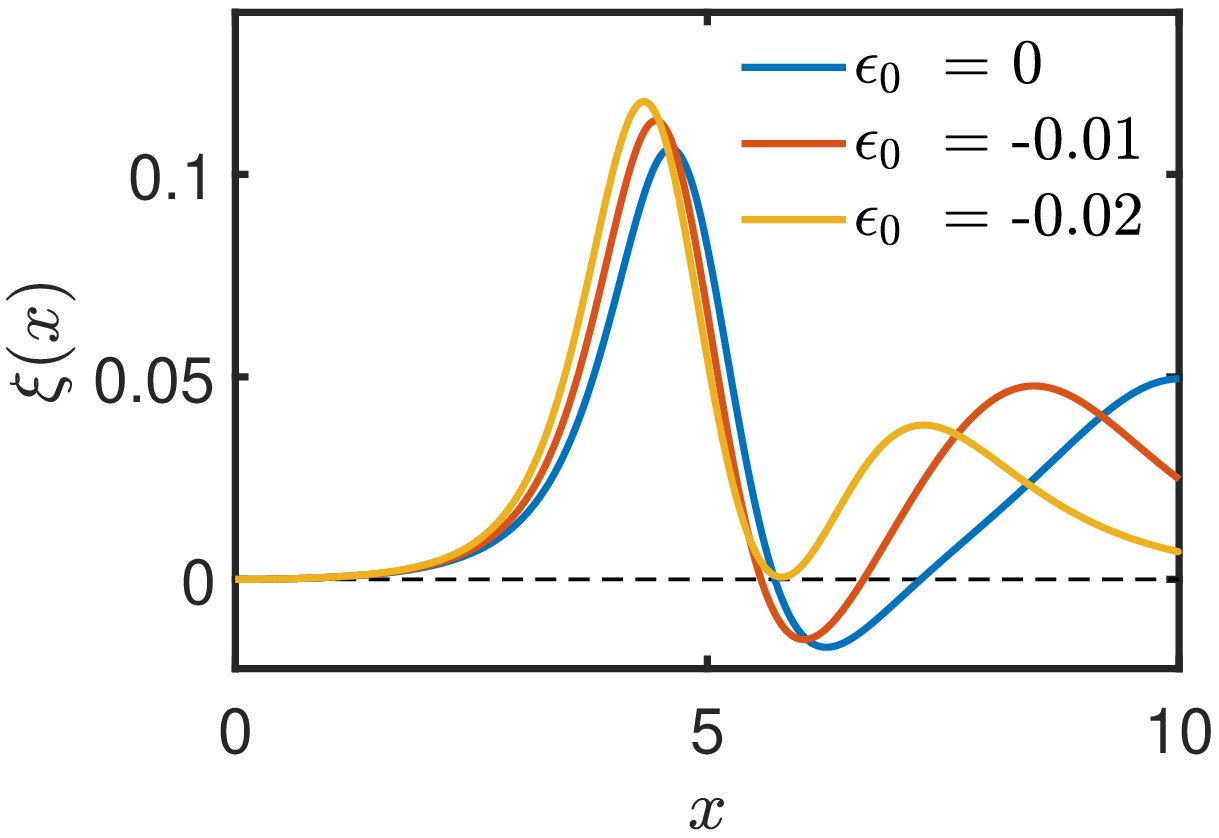}
\caption{}
\label{1c}
\end{subfigure}

\caption{(a) Diffusion coefficient $D(x)$ and  (b) viscosity $\xi (x)$ as functions of nuclear position computed for different values of the applied voltage and with resonant energy level set to zero $\epsilon_0=0$; (c) viscosity $\xi (x)$ as a function of bond-length computed for different resonant-level energies $\epsilon_0$ at voltage $V = 0.06$.    }

\end{figure*}

Each of our calculations utilize a common set of unchanging parameters: the bandwidth for numerical integration is set to [-5, 5]; left and right lead temperatures are set to be equal ($T_L = T_R = 300K$); the reduced mass associated with the chemical bond is $m=1000$; the molecule is always strongly coupled to the left lead with $\Gamma_L = 4$     and $\Gamma_R = 0.03$; and the equilibrium bond-length is $x_0 = 5$. All numerical values  in the text and figures are given in atomic units.

\subsubsection{Electronic friction, diffusion coefficient, and local effective temperature}

We first study how  the parameters of the model control the three main ingredients of the Langevin equation:  the diffusion coefficient, viscosity, and adiabatic force. Fig. \ref{1a} shows the diffusion coefficient $D(x)$ as a function of the bond-length. The amplitude of the random force is the square root of the diffusion coefficient. As seen in Fig.\ref{1a}, the diffusion coefficient has a strong dependence on the bond-length, reaching its maximum at the equilibrium bond-length and then decaying to zero as the bond stretches or contracts. As physically expected, the amplitude of the random force increases as the voltage becomes larger.

The viscosity $\xi(x)$ is shown in Fig.\ref{1b}. At small voltages the viscosity behavior mirrors the diffusion coefficient's dependence on the bond-length. This is not surprising if one recalls the fluctuation-dissipation theorem which relates the ratio of the diffusion coefficient  $D(x)$ and  viscosity $\xi (x)$ to the temperature, and temperature should not deviate significantly from the equilibrium value for small voltages.  If the voltage is increased, we start to observe regions of negative viscosity  which energize the stretching/contraction of the bond rather than dampening its oscillations  as one may expect from the viscous force. This negative viscosity phenomenon has been previously observed for similar theoretical systems using varying modeling techniques\cite{Bode12,Lu11,Metelmann11}.

Fig. \ref{1c} shows viscosity as a function of bond-length computed at $V=0.06$     of applied voltage. Once the level moves away from the resonance position $\epsilon_0 =0$, the second peak in the viscosity starts to shift closer to the equilibrium bond-length. The second peak occurs when the energy of the level intersects the Fermi level of the right lead, such that electrons are easily able to transition between the lead and the resonance level, while the left lead is essentially disconnected due to the exponential coupling decay.

\begin{figure*}
\begin{subfigure}{0.4\textwidth}
\centering
\includegraphics[width=1\textwidth]{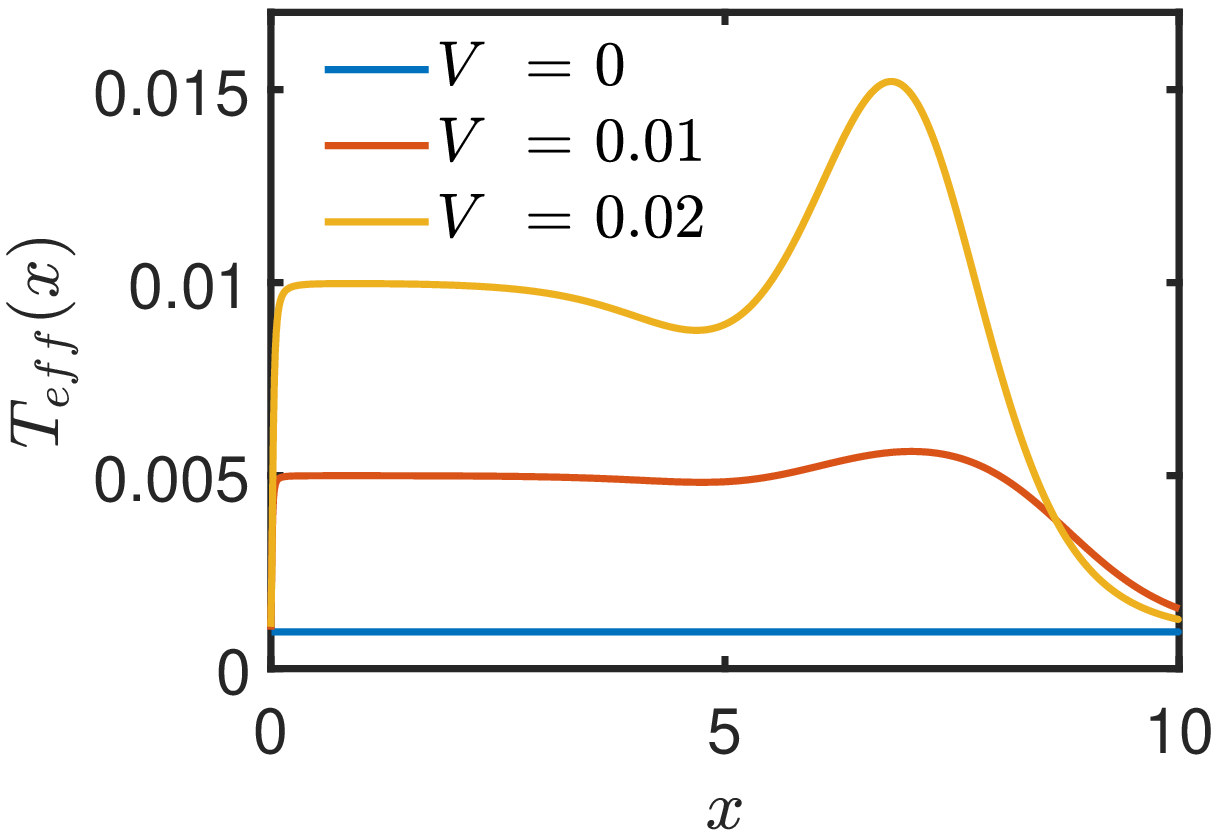}
\caption{}
\label{2a}
\end{subfigure}
\begin{subfigure}{0.4\textwidth}
\centering
\includegraphics[width=1\textwidth]{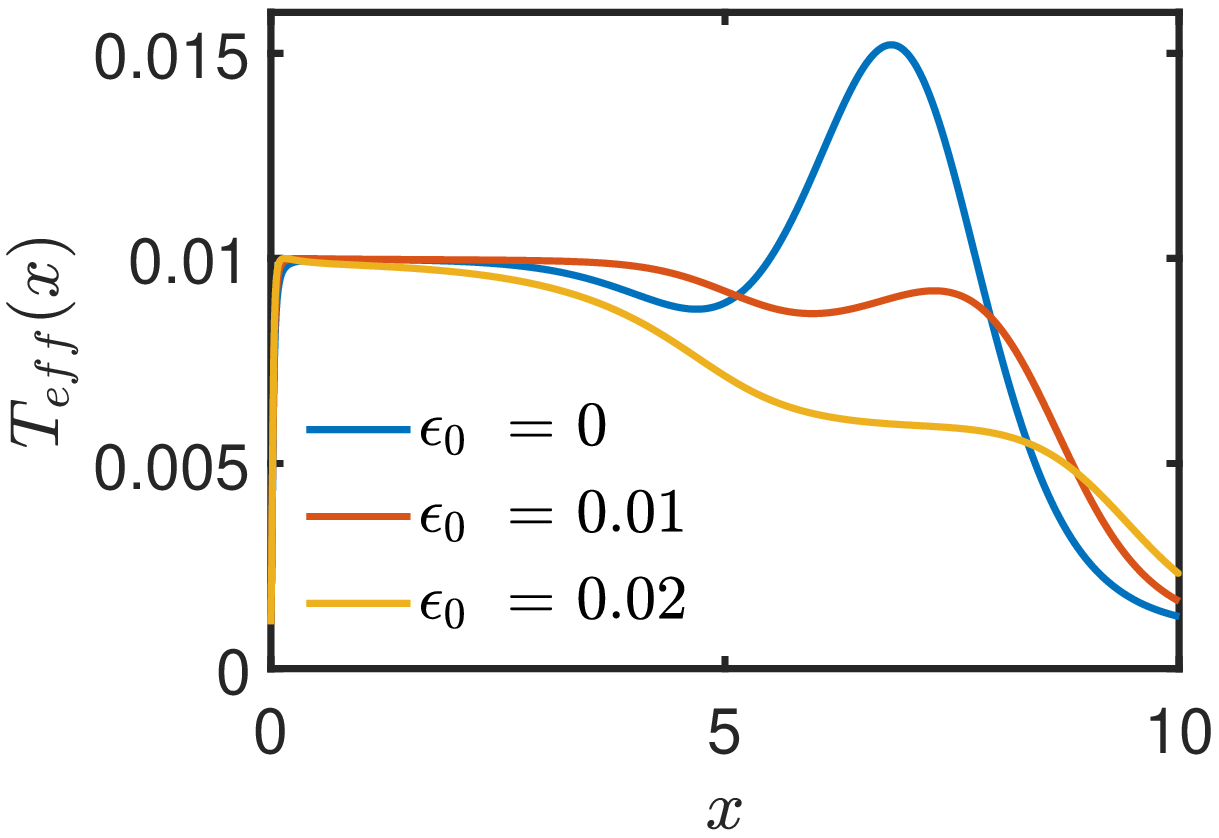}
\caption{}
\label{2b}
\end{subfigure}
\begin{subfigure}{0.4\textwidth}
\centering
\includegraphics[width=1\textwidth]{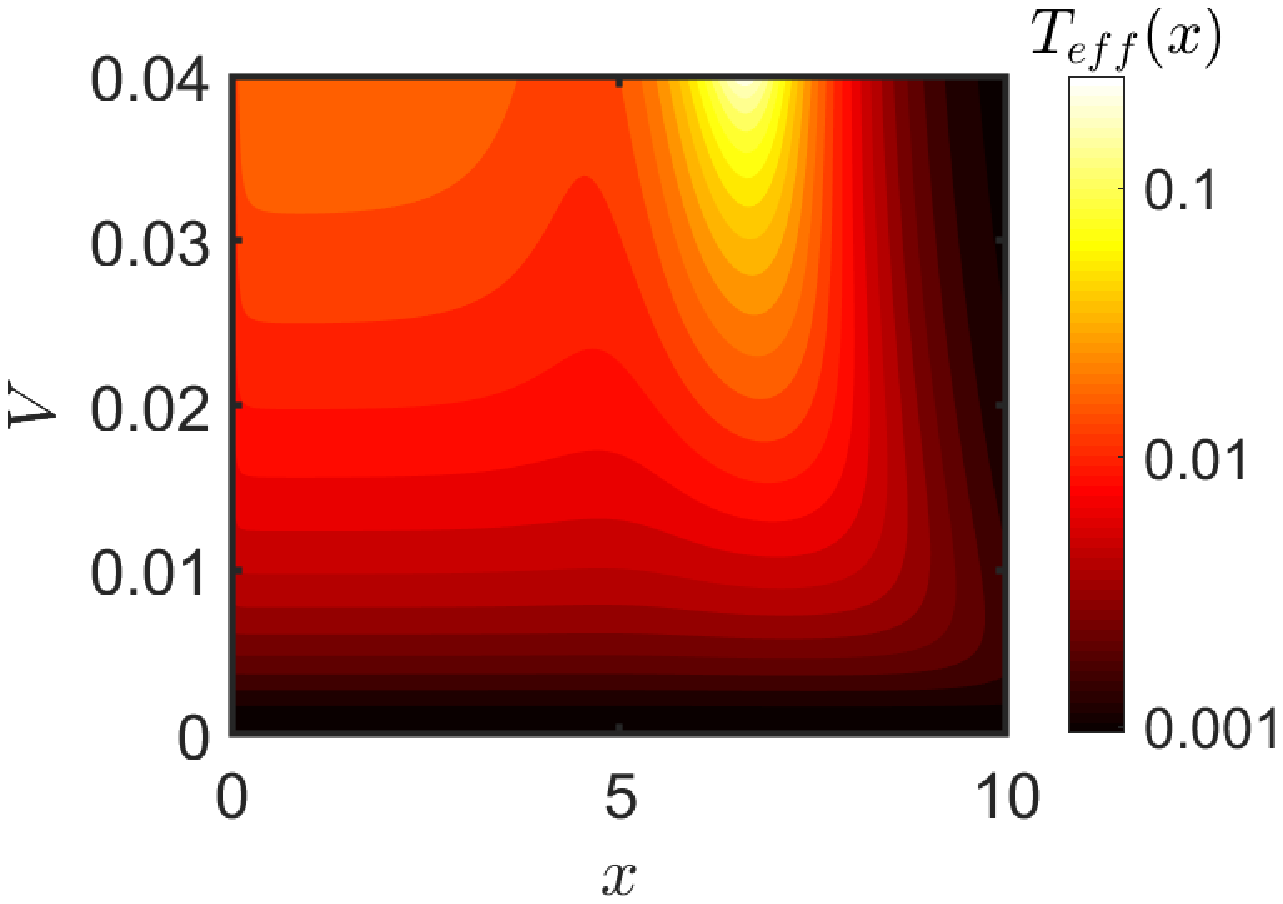}
\caption{}
\label{2c}
\end{subfigure}
\begin{subfigure}{0.4\textwidth}
\centering
\includegraphics[width=1\textwidth]{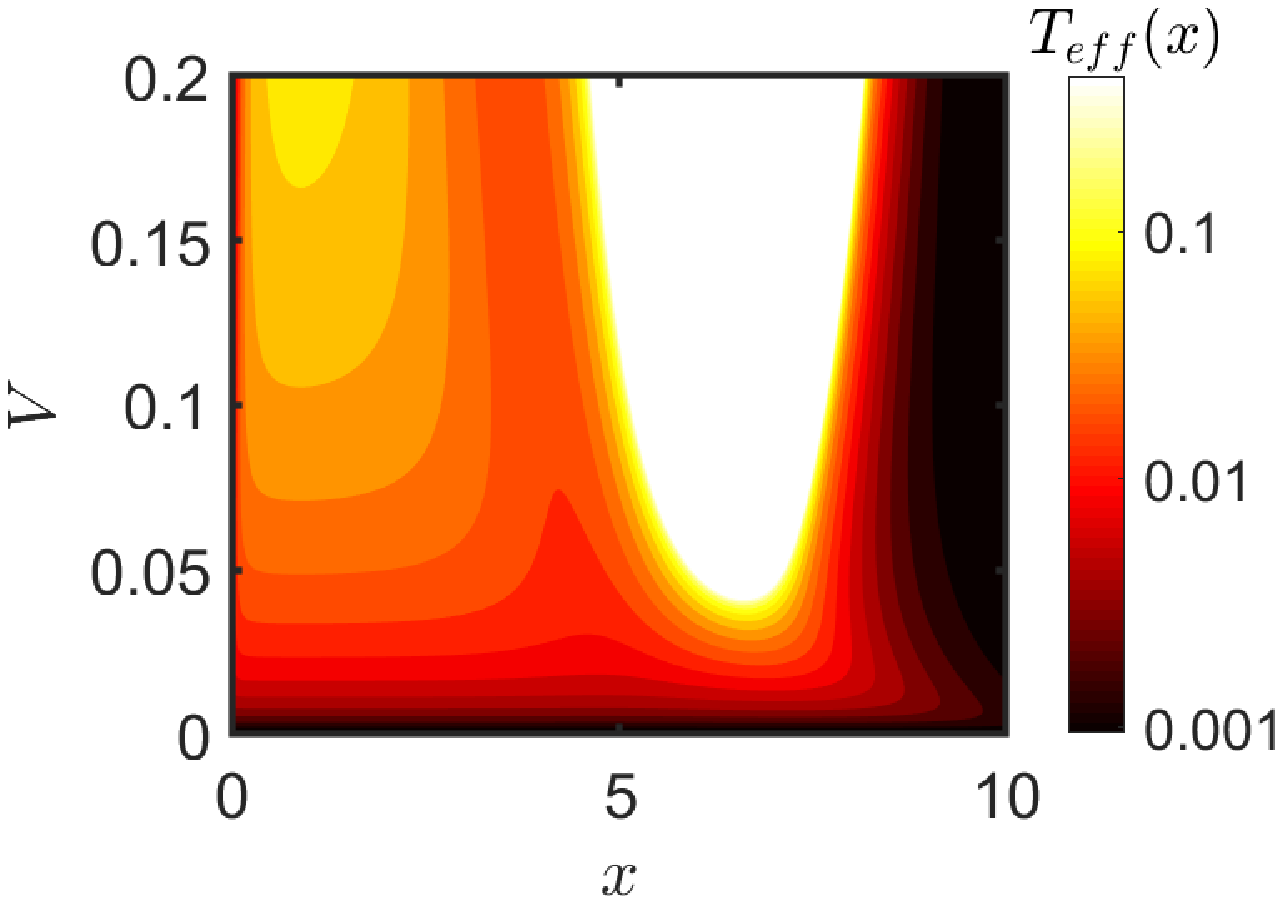}
\caption{}
\label{2d}
\end{subfigure}

\caption{(a) Effective temperature $T_{\text{eff}}(x)$ as a function of bond-length computed in the resonance regime $\epsilon_0=0$  at various values of the applied voltage. (b)  Effective temperature $T_{\text{eff}}(x)$ as a function of bond-length computed at applied voltage  $V=0.02$      for different values of the resonant-level energies. (c) Contour plot of effective temperature  $T_{\text{eff}}(x)$  as a function of voltage and bond-length  for low voltages.(d) Contour plot of effective temperature  $T_{\text{eff}}(x)$  as a function of voltage and bond-length for high voltages; the white region represents negative effective temperatures. }
\end{figure*}

In analogy to the fluctuation-dissipation theorem \cite{zwanzig-book}, it is instructive to define an effective temperature as 
\begin{equation}
T_{\text{eff}}(x) = \frac{D(x)}{2\xi (x)}.
\end{equation}
This effective temperature is an intuitively clear physical quantity  which reveals information on the steady-state spatial distribution of kinetic energy within the junction and is related to current-induced localized heating or cooling effects.

\begin{figure}
\centering
\begin{subfigure}{0.5\textwidth}
\centering
\includegraphics[width=1\textwidth]{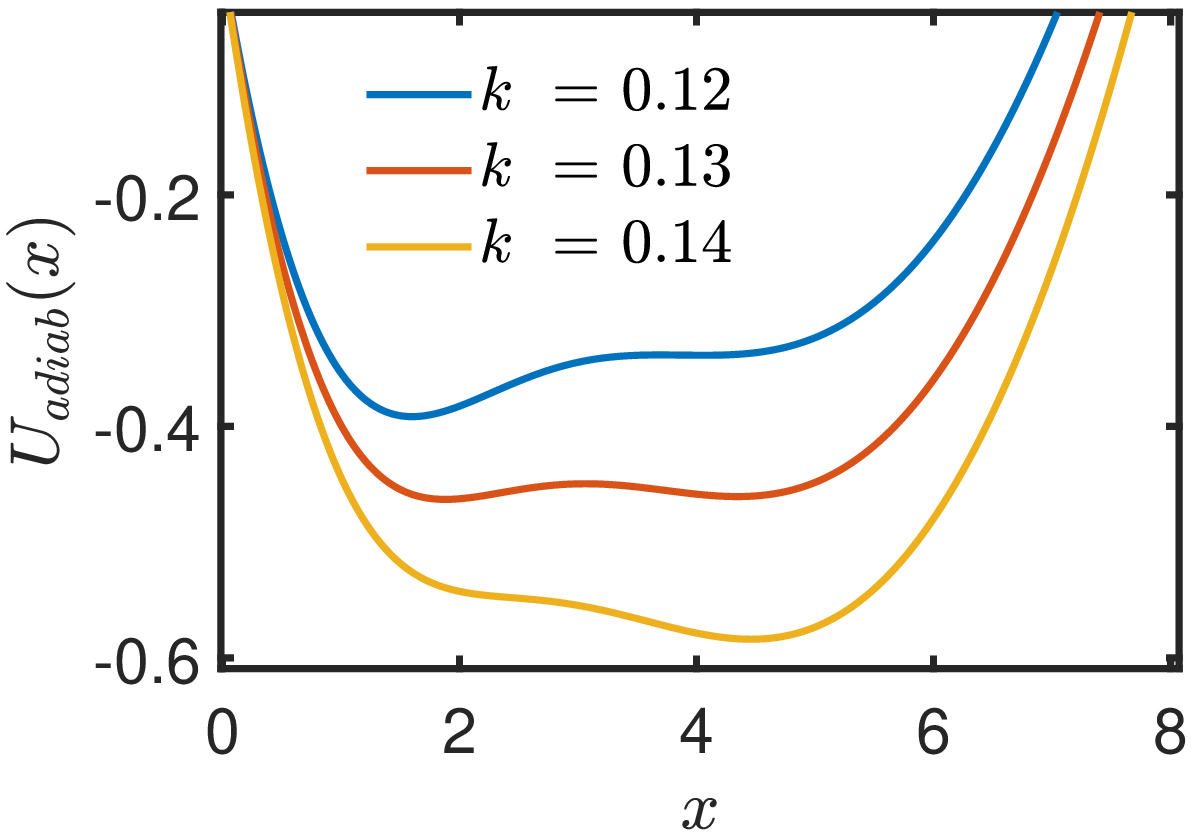}
\caption{}
\label{3a}
\end{subfigure}
\begin{subfigure}{0.5\textwidth}
\centering
\includegraphics[width=1\textwidth]{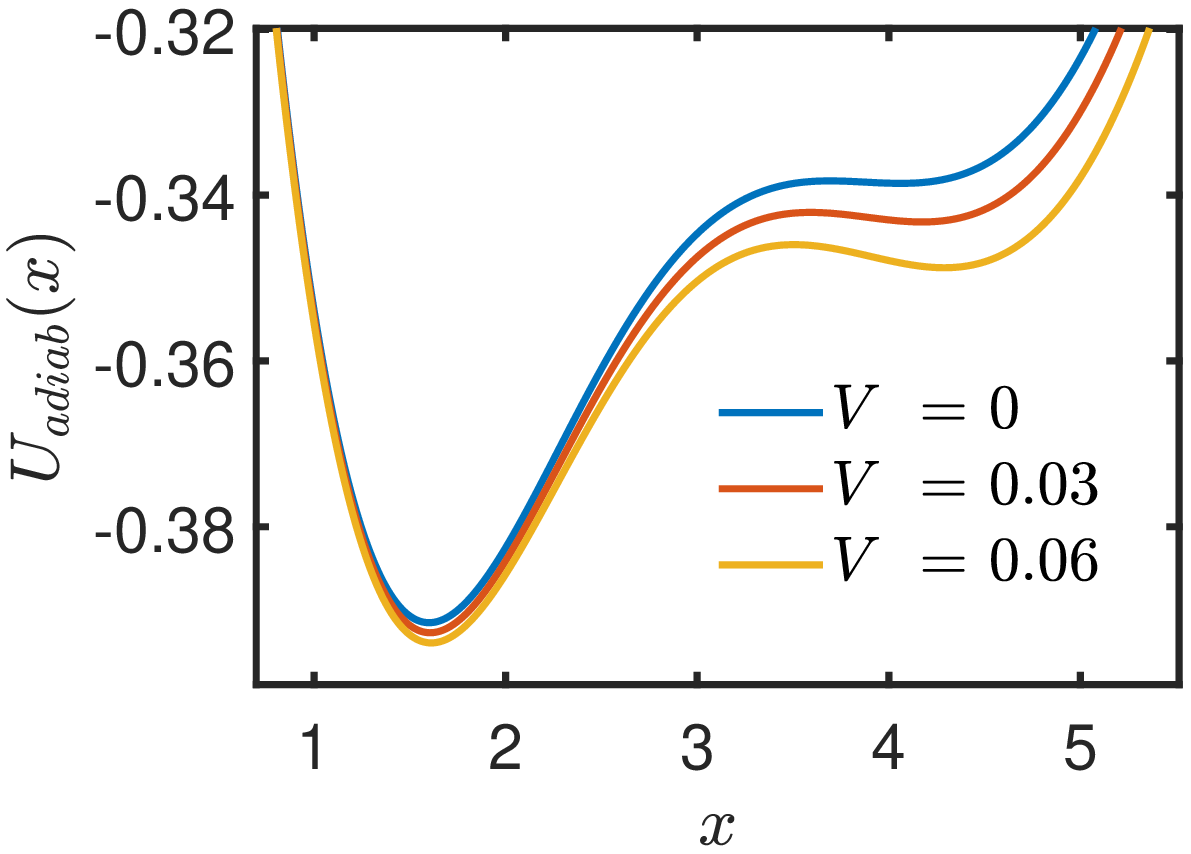}
\caption{}
\label{3b}
\end{subfigure}

\caption{Adiabatic potential for (a) varying spring constants ($V = 0$), and  (b)  for varying bias voltages ($k = 0.12$). }

\end{figure}

\begin{figure}

\centering
\begin{subfigure}{0.5\textwidth}
\centering
\includegraphics[width=1\textwidth]{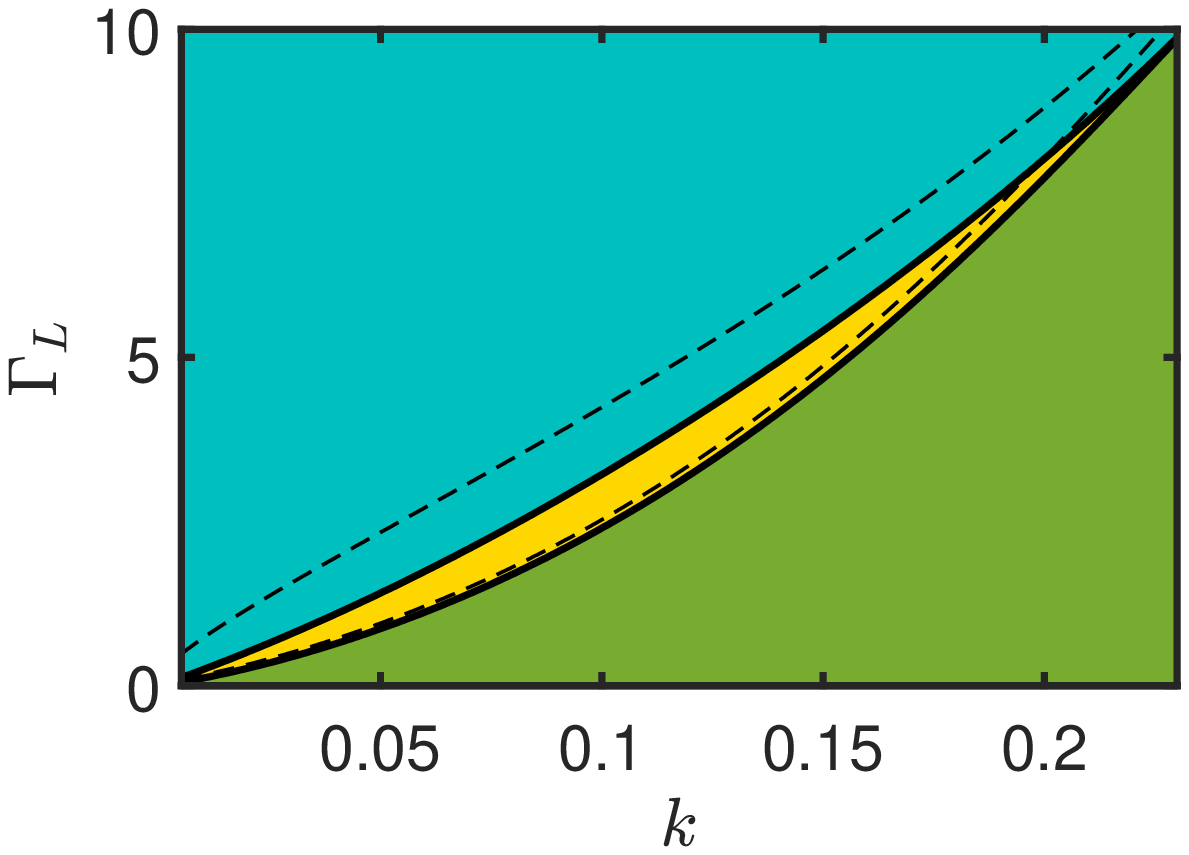}
\caption{}
\label{fig4a}
\end{subfigure}
\begin{subfigure}{0.26\columnwidth}
\centering
\includegraphics[width=1\columnwidth]{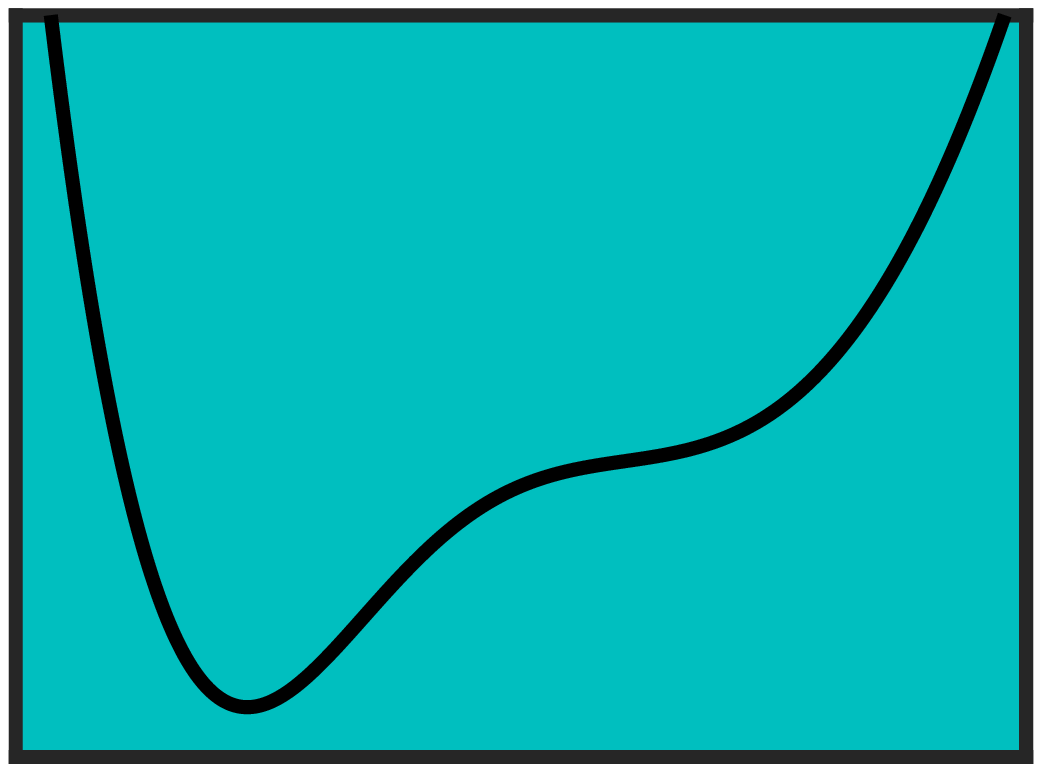}
\caption{}
\label{fig4b}
\end{subfigure}
\begin{subfigure}{0.26\columnwidth}
\centering
\includegraphics[width=1\columnwidth]{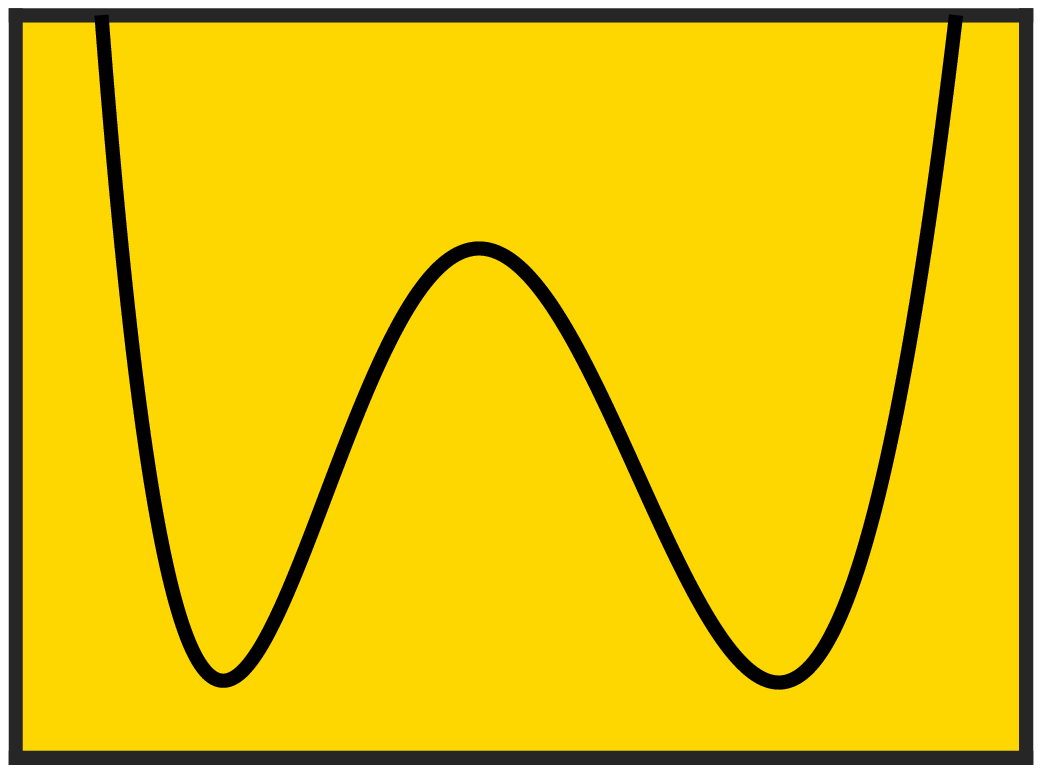}
\caption{}
\label{fig4c}
\end{subfigure}
\begin{subfigure}{0.26\columnwidth}
\centering
\includegraphics[width=1\columnwidth]{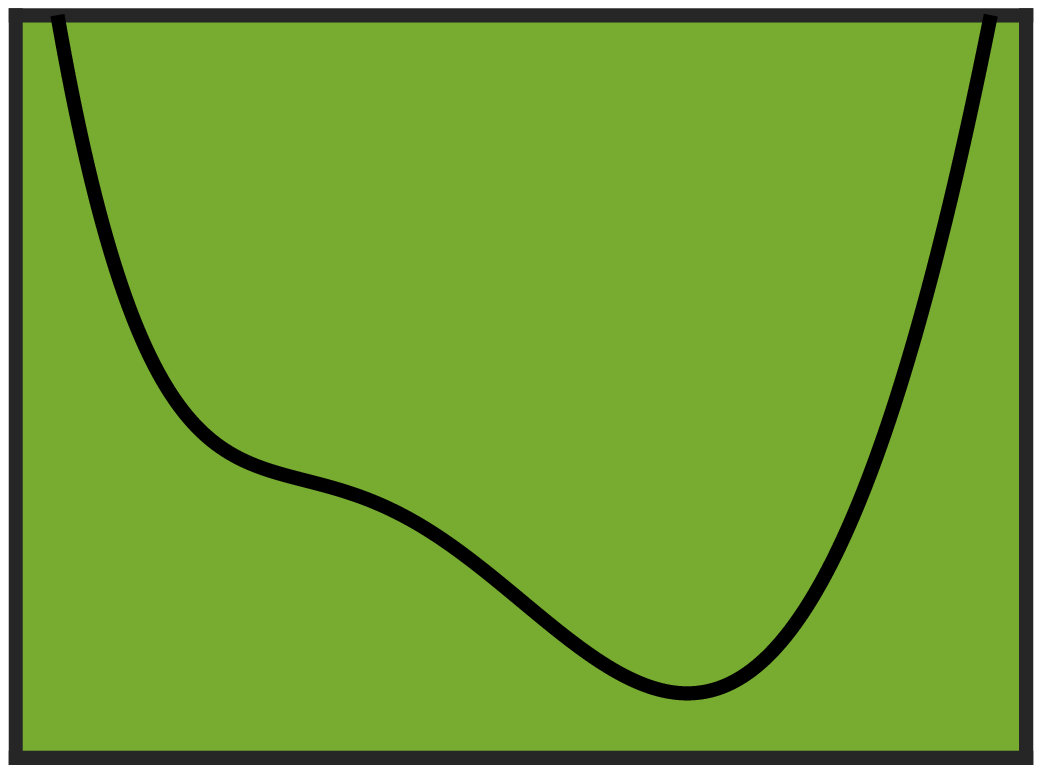}
\caption{}
\label{fig4d}
\end{subfigure}

\caption{(a) Different adiabatic potential regimes for varying $k$ and $\Gamma_L$,  computed at $V=0$ and color coded according to (b), (c), (d).  Dashed lines show the boundaries of the bistable yellow region for $V=0.2$.}
\end{figure}

It is clear from Fig.\ref{2a} that in the equilibrium case (zero applied voltage), the fluctuation-dissipation theorem is satisfied as $T_{\text{eff}}$ is independent of $x$ and equals to $300$ K, exactly the temperature of left and right leads. Once the  voltage is increased, the current carrying electrons produce significant local heating in the junction leading to the rise of the effective temperature. The coordinate dependence of effective temperature has a small dip at equilibrium bond-length and then reaches its maximum value if the bond is stretched.

 In Fig.\ref{2d}, we observe a region of parameters in our junction in which the effective temperature becomes negative, such that the nucleus has no defined steady-state local kinetic energy in this region and as such, the kinetic energy of the nuclei will continue to increase if constrained to this region.

Next, we compute the adiabatic potential as a function of bond-length. By combining the classical  potential and integrating our adiabatic force $F_{(0)}(x)$ computed by Eq.\ref{f0model}, we obtain the adiabatic potential \cite{dzhioev11}
\begin{equation}
{U_{\text{adiab}}}(x) = U(x) - \int ^x_{a} dx' F_{(0)}(x').
\end{equation}
Notice that the lower limit in this integral $a$ is completely arbitrary and serves as a reference point for the computed potential energy.  We use $a=0$     in all our calculations.
We observe in \ref{3a} and \ref{3b} the possibility of different potential regimes in which we may observe two separate stable minima. 

These regimes are summarized according to the changing bond spring constant and coupling  in Figure 4. There is a narrow region of bistability.
Once we move away from this region, one minimum starts to dominate  until the other minimum disappears completely.  As one increases the voltage, the bistable yellow region becomes wider and shifts towards smaller values of the spring constant.

\begin{figure*}
\centering
\begin{subfigure}{0.4\textwidth}
\centering
\includegraphics[width=1\textwidth]{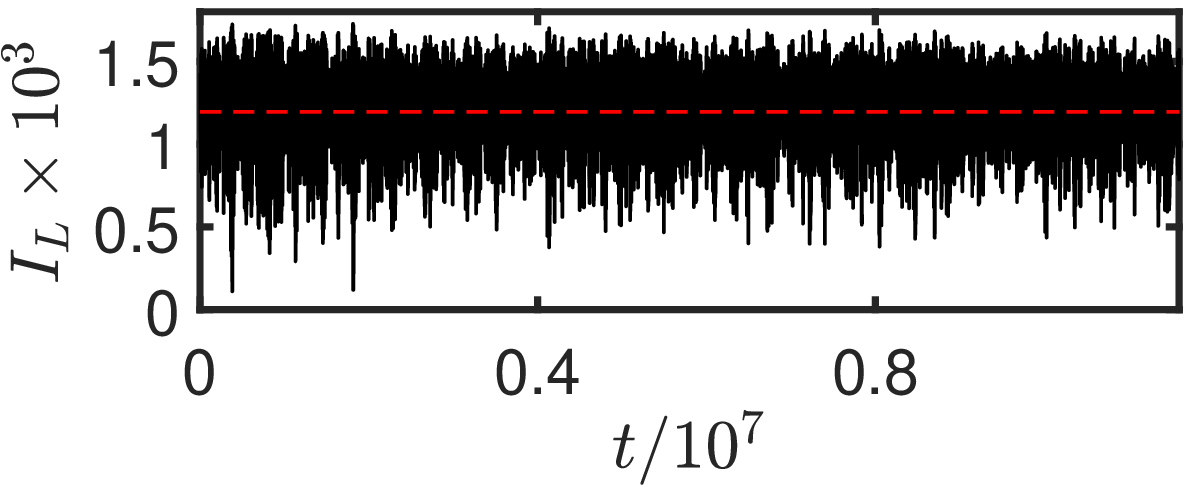}
\caption{}
\label{5a}
\end{subfigure}
\begin{subfigure}{0.4\textwidth}
\centering
\includegraphics[width=1\textwidth]{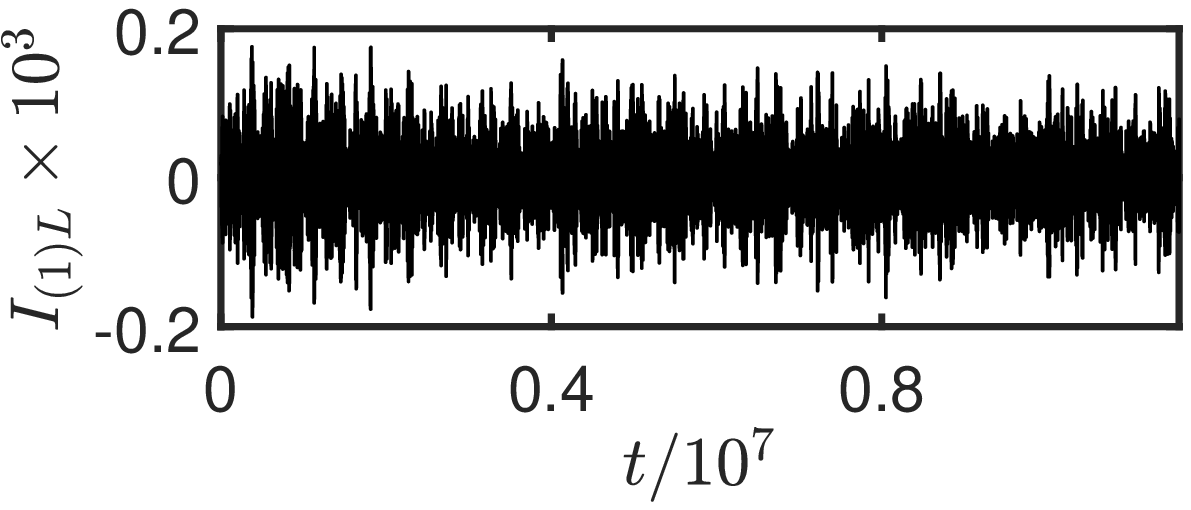}
\caption{}
\label{5b}
\end{subfigure}

\begin{subfigure}{0.4\textwidth}
\centering
\includegraphics[width=1\textwidth]{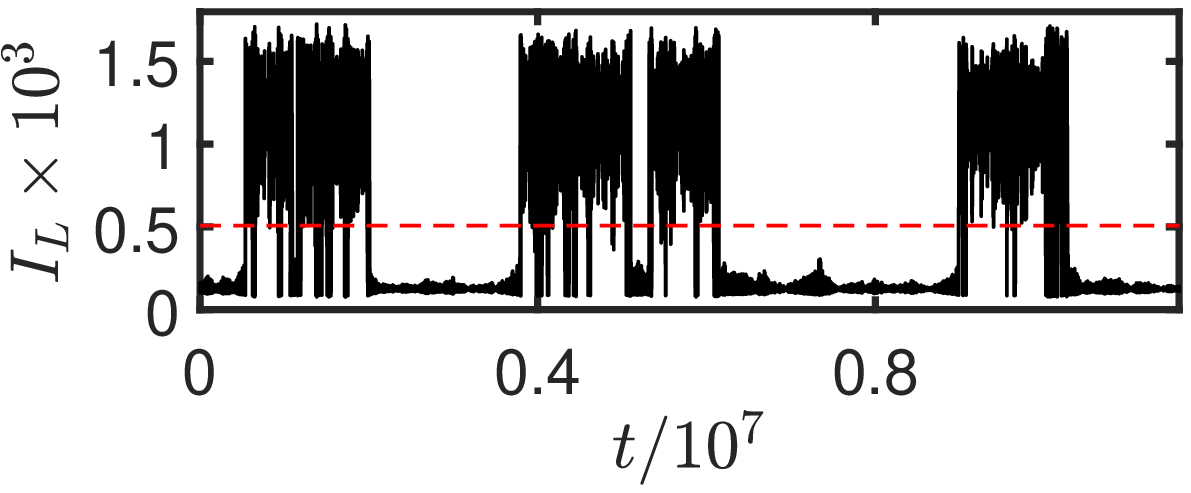}
\caption{}
\label{5c}
\end{subfigure}
\begin{subfigure}{0.4\textwidth}
\centering
\includegraphics[width=1\textwidth]{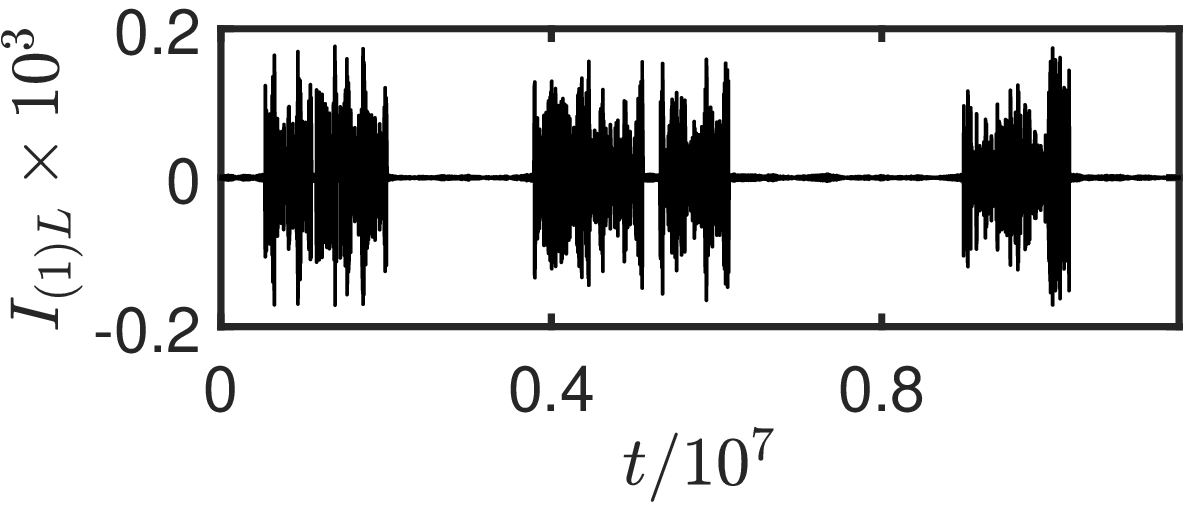}
\caption{}
\label{5d}
\end{subfigure}

\begin{subfigure}{0.4\textwidth}
\centering
\includegraphics[width=1\textwidth]{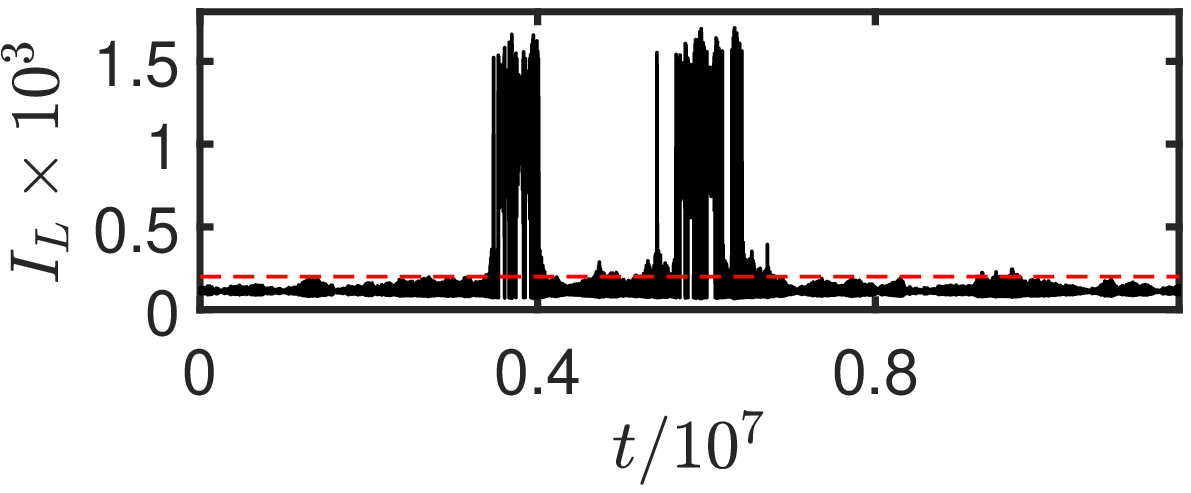}
\caption{}
\label{5e}
\end{subfigure}
\begin{subfigure}{0.4\textwidth}
\centering
\includegraphics[width=1\textwidth]{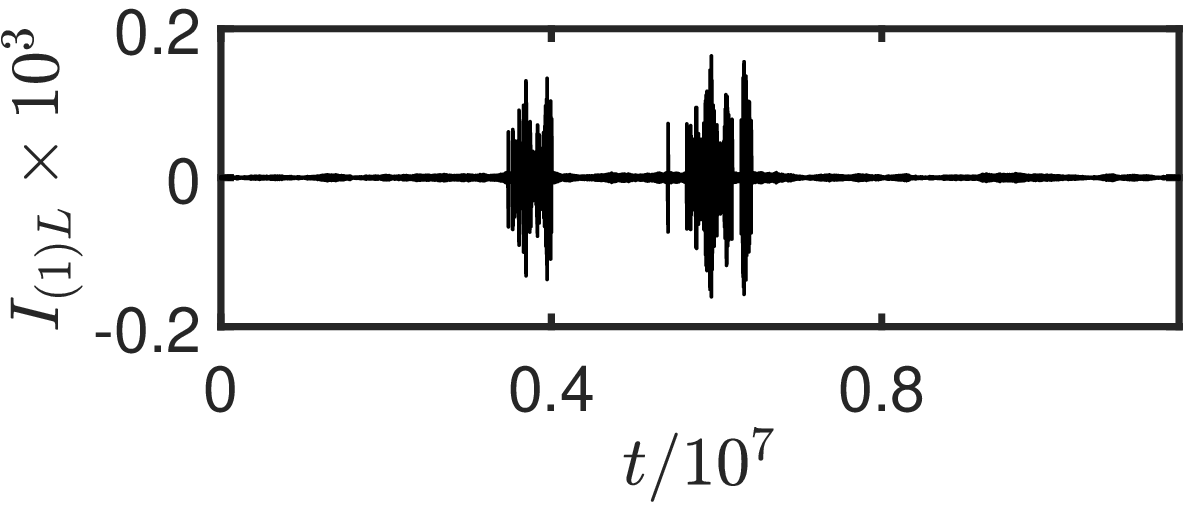}
\caption{}
\label{5f}
\end{subfigure}

\caption{Current with dynamical corrections $I_L(t)= I_{L}^{(0)}(t)+ I_{L}^{(1)}(t) $ and first order correction $I_{L}^{(1)}(t) $ to the current as functions of time  computed at $V=0.01$ a.u  for different values of the spring constant (a,b) $k = 0.136$  (c,d) $k = 0.131$ (e,f) $k = 0.127$. The red dashed line denotes the current mean over the displayed time interval.}
\end{figure*}

\subsubsection{Current}

In this section, we show results for the current computed along a given trajectory of the bond-length time-evolution obtained from the solution of the Langevin equation. 
To compute a trajectory  $x(t)$, we utilize an m-BAOAB algorithm provided by Sachs et al\cite{algorithm}, which enables a numerical  solution of the Langevin equation with a coordinate dependent viscosity and diffusion coefficient. The trajectory is used to compute Green's functions, and current with first order dynamical corrections using the equation presented in section \ref{model}.
We  consider three representative scenarios with very distinct nuclear dynamics: rigid chemical bonding ($k=0.136$), intermediate chemical bonding  ($k = 0.131$), and soft chemical bonding ($k = 0.127$). 
In the case of a rigid chemical  bond, the bond-length oscillates around a single minima; this is reflected in the time dependence of current shown in Fig.\ref{5a}, \ref{5b}. 
Both the electric current with dynamical corrections $I_L(t)= I_{L}^{(0)}(t)+ I_{L}^{(1)}(t) $ and the first order correction $I_{L}^{(1)}(t) $ itself oscillate around single average values. 
Once the chemical bond becomes softer ($k = 0.131$),  the length of the chemical bond switches between two states, spending roughly equal time in each. This behavior of the bond-length results into telegraphic switching of the current between two values as shown in Fig.\ref{5c}, \ref{5d}. The first order dynamical correction is more noticeable in the more conducting state. For a soft molecule-lead chemical bond, $k=0.127$, the bond-length experiences switching  but has a preference for a specific value, as does the current.

\subsection{ Current noise}
The temporal correlations between stochastic fluctuations of the electric
current (current noise) have become a very important experimental
and theoretical tool in studying transport properties of molecular
junctions. Noise spectroscopy enables the study of the special features
of a single-molecule junction, which are not accessible by standard
current-voltage measurements. The experimental noise measurements
provide significantly new information on fundamental mechanisms of
electron transport in molecular junctions, such as atomistic details
of the local environment and metal-molecule interfaces\cite{noise10,doi:10.1021/acs.nanolett.5b01270},
coupling between electronic and vibrational degrees of freedom\cite{galperin06a,PhysRevLett.106.136807,PhysRevLett.108.146602,thoss14},
identifications of the individual conduction transport channel\cite{doi:10.1021/nl060116e,doi:10.1021/nl904052r,Tsutsui:2010aa,noise17},
and mechanical stability of the junction\cite{C4NR03480E}.

\begin{figure*}
\begin{subfigure}{0.45\textwidth}
\centering
\includegraphics[width=1\textwidth]{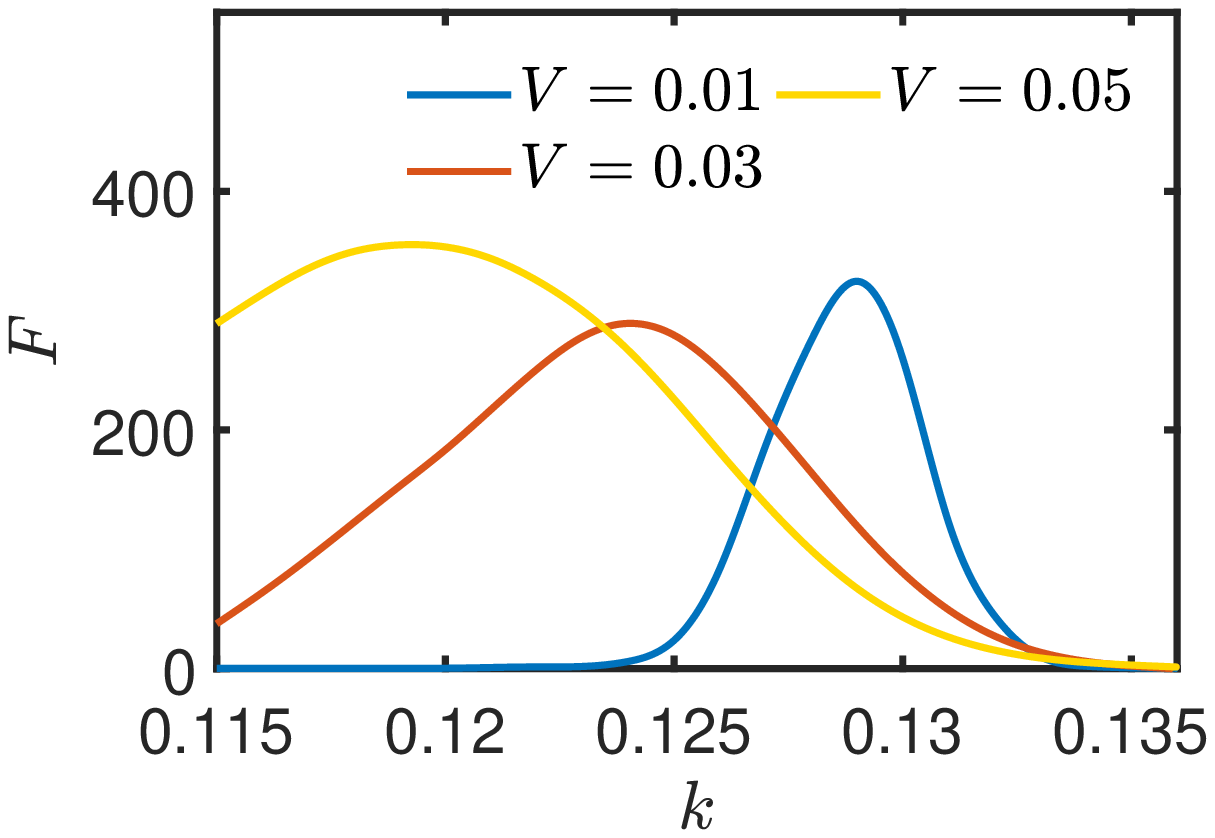}
\caption{}
\label{6a}
\end{subfigure}
\begin{subfigure}{0.45\textwidth}
\centering
\includegraphics[width=1\textwidth]{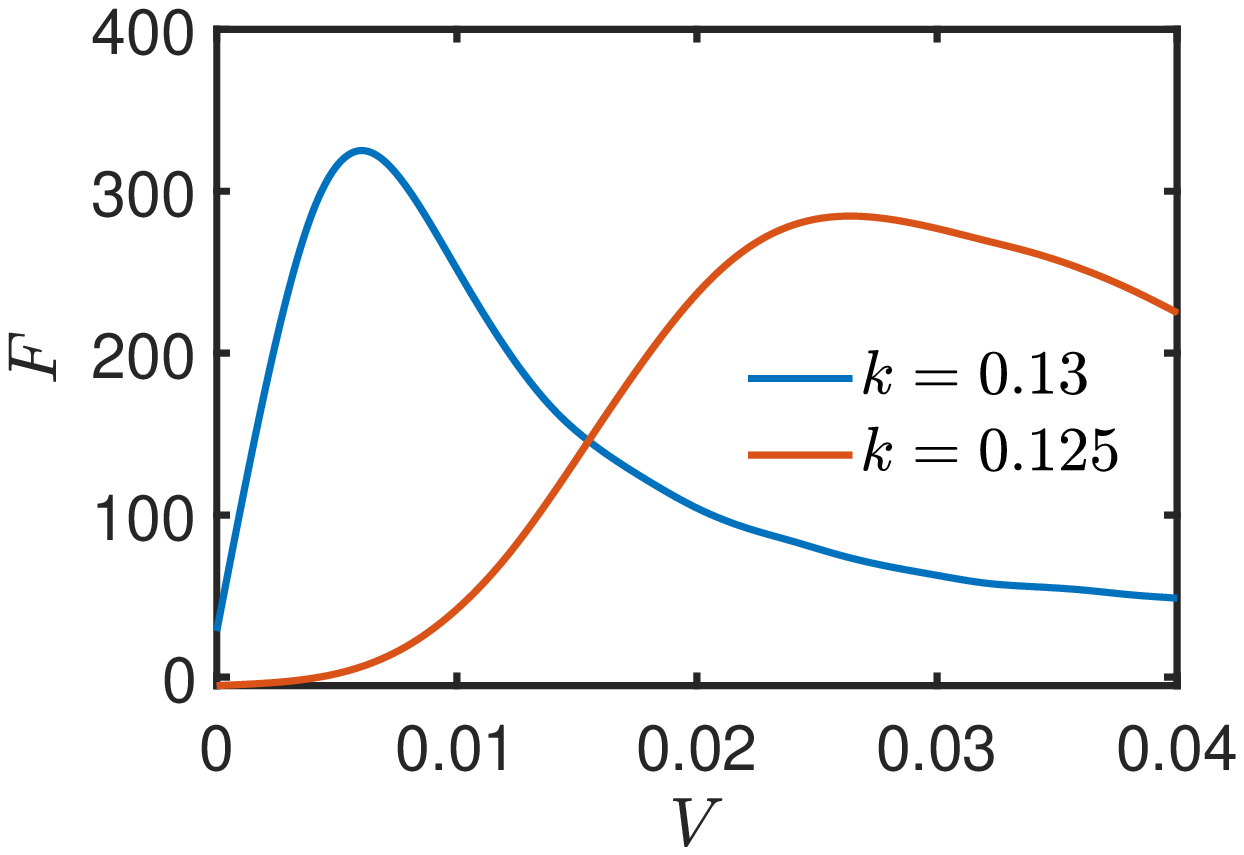}
\caption{}
\label{6b}
\end{subfigure}

\begin{subfigure}{0.45\textwidth}
\centering
\includegraphics[width=1\textwidth]{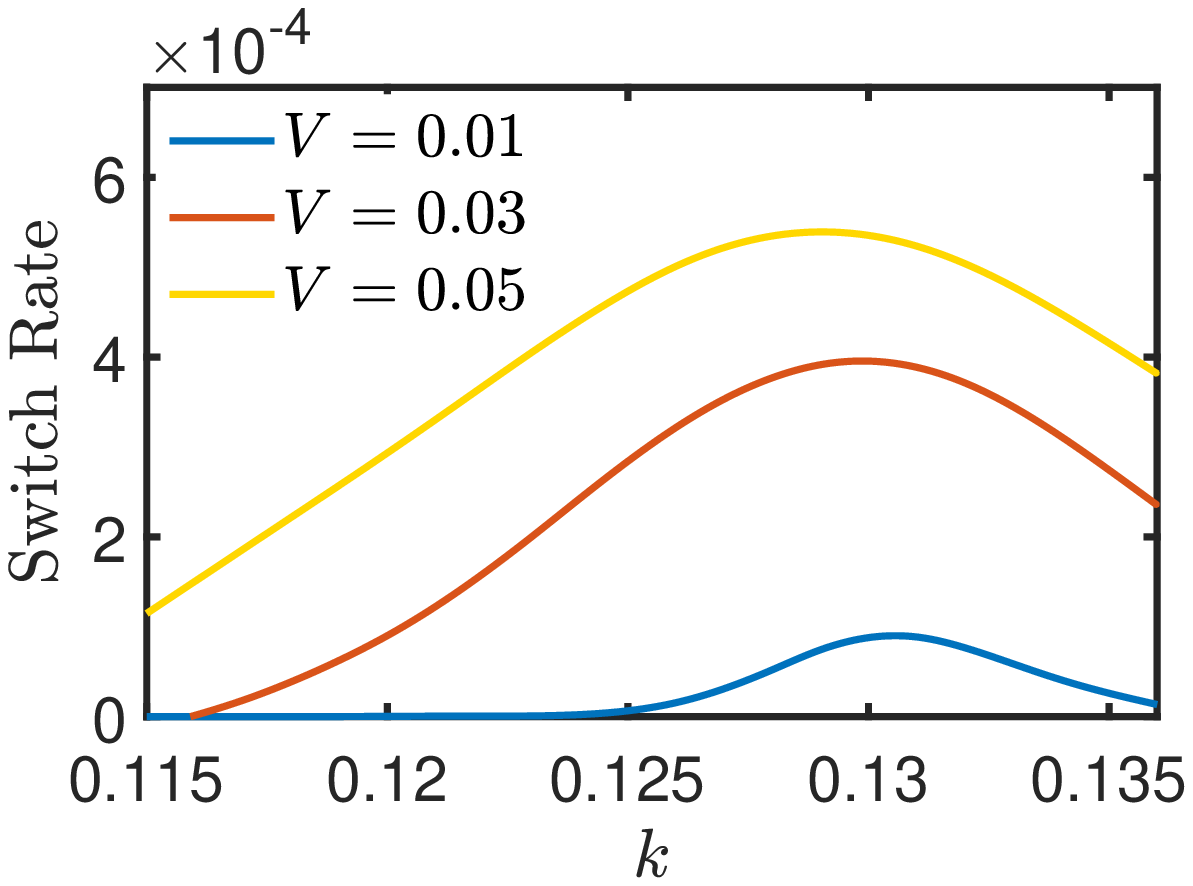}
\caption{}
\label{6c}
\end{subfigure}
\begin{subfigure}{0.45\textwidth}
\centering
\includegraphics[width=1\textwidth]{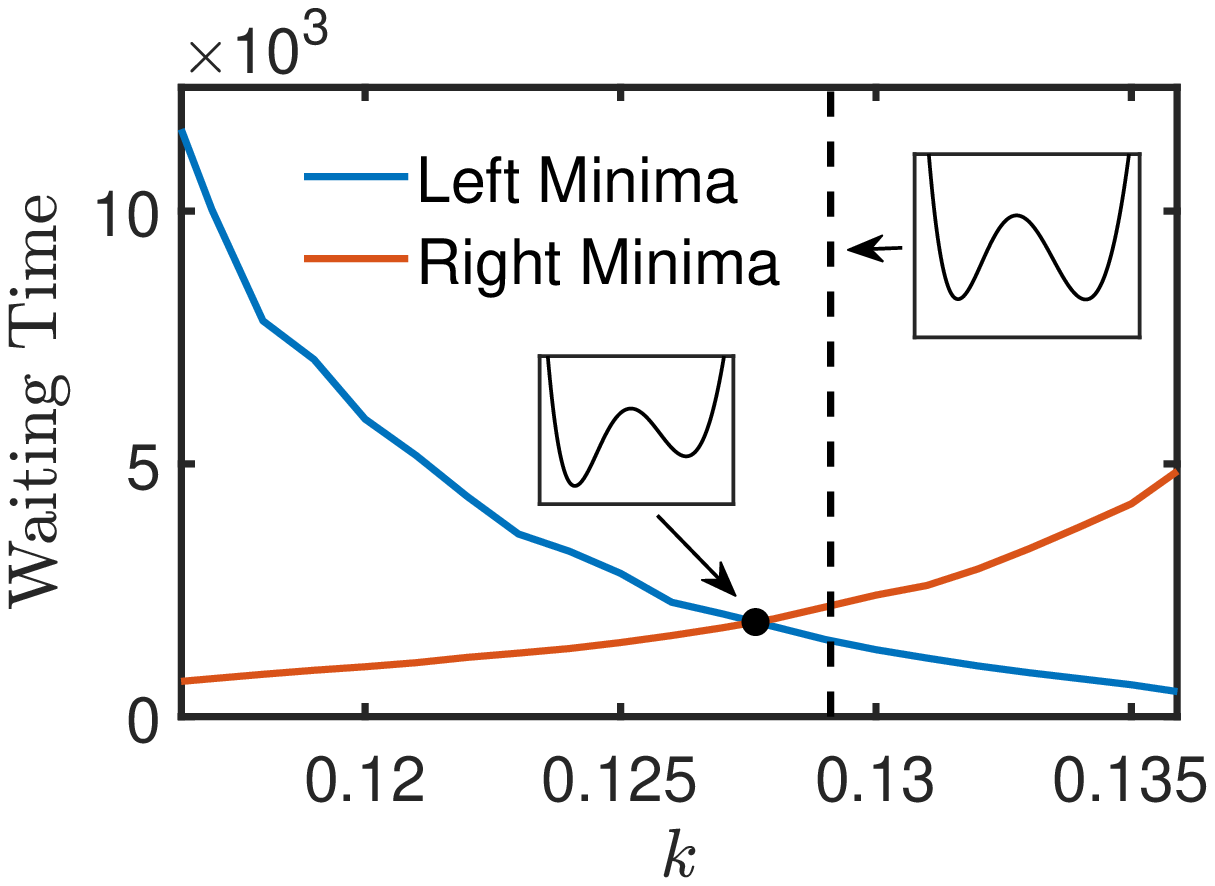}
\caption{}
\label{6d}
\end{subfigure}

\label{FIG. 6}

\caption{(a) Fano factors as  functions of the spring constant $k$. (b) Fano factors as  functions of voltage $V$. Here $k=0.13$ yields two minima with approximately equal depth, while $k=0.125$ yields a deeper left minimum. (c) Average switch rate between minima in a bi-stable regime, varying $k$. (d) Average waiting times in a bi-stable regime for $V = 0.05$. The vertical dashed line denotes the $k$ value for which the two minima have equal depth.  All calculations are performed for  $\epsilon_0 = 0$.}
\end{figure*}

Current noise is  formally defined as 
\begin{equation}
S_{\alpha}(\tau)=\lim_{T\rightarrow+\infty}\frac{1}{T}\int _{0}^{T}dt\langle\Big[\delta\hat{I}_{\alpha}(t),\delta\hat{I}_{\alpha}(t+\tau)\Big]_{+}\rangle,\label{s1}
\end{equation}
where $\delta\hat{I}(t)$ describes the instantaneous deviation
of the electric current at time $t$ from its average value and $[...,...]_{+}$
is the anti-commutator. Eq.(\ref{s1}) involves two averages: $\langle...\rangle$
is the quantum expectation value over electronic degrees of freedom and
$\lim_{T\rightarrow+\infty}\frac{1}{T}\int _{0}^{T}dt...$ is the time
average over the classical motion of the nuclei. The time average is equivalent
to the ensemble average over many realizations of geometries of the
molecular junction. The current noise power spectrum is the Fourier
transformation of (\ref{s1}) 
\begin{equation}
S_{\alpha}(\omega)=\int ^{+\infty}_{-\infty} d\tau e^{i\omega\tau}S_{\alpha}(\tau).
\end{equation}
The electric current noise provides valuable information about the system and originates from multiple factors:
(a) the quantum nature of electrons, discreteness of charge, Pauli exclusion principle, shot noise, and the finite temperature of electrons; (b) various types of quantum correlations between current-carrying electrons, which are not present in our model; (c) and finally, the "mechanical" noise due to current-induced changes to the molecular junction geometry. Generally the total noise is not simply the addition of (a), (b), and (c) contributions; there is a cross interference between different contributions. However, within our approach the distinctly different time-scales of fast electronic and slow nuclear
motion enables the separation of  the mechanical noise contribution\cite{PhysRevB.83.035420}.
 The characteristic time scale of shot noise decay is $1/\Gamma$,   whereas the noise due to nuclear motion appears on much longer times. Hence the noise induced by geometrical fluctuations dominates the noise power spectrum at low frequencies, and can exceed the shot noise contribution by orders of magnitude \cite{PhysRevB.83.035420}.

In what follows we focus on the "mechanical" noise as 
\begin{equation}
S_{\alpha}(\tau)=2\lim_{T\rightarrow+\infty}\frac{1}{T}\int _{0}^{T}dt\delta I_{\alpha}(t)\delta I_{\alpha}(t+\tau),
\label{currentnoise}
\end{equation}
where the current fluctuation at time $t$ is 
\begin{equation}
\delta I_{\alpha}(t)=I_{\alpha}(t)-\lim_{T\rightarrow+\infty}\frac{1}{T}\int _{0}^{T}dtI_{\alpha}(t).
\end{equation}

The Fano factor is 
\begin{equation}
F_{\alpha}=\frac{S_{\alpha}(\omega=0)}{2I_{\alpha}}.
\label{fanofactor}
\end{equation}
The variance and mean of a Poisson process is equal, therefore the Fano factor can be used  to characterize electron transport as either a sub-Poissonian ($F < 1$), Poissonian ($F = 1$), or super-Poissonian ($F > 1$) process. Indeed, super-Poissonian or sub-Poissonian noise is caused by a host of very interesting and often hidden physical effects.
\begin{figure*}
\centering
\includegraphics[width=1 \textwidth]{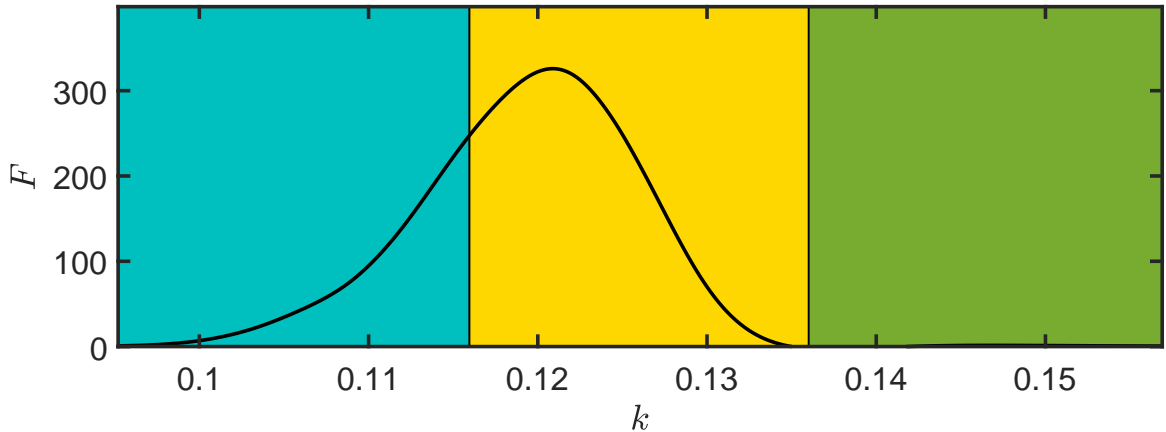}
\caption{Fano factor computed as a function of the spring constant; $V=0.05$ and $\epsilon_0=0$.  The background of the figure is color-coded in accordance to the different regimes of adiabatic potential shown in Fig.4 (b,c,d).}
\label{FIG. 7}
\end{figure*}

Figs.\ref{6a} and \ref{6b} show Fano factors computed as a functions of the applied voltage $V$ and spring constant $k$.
The presence of telegraphic switching between two minima in a bi-stable adiabatic potential results in 
the gigantic enhancement of the Fano factor, indicating that the electron transport is a super-Poissonian process. 
The behavior of the  Fano factor depends on a number of  factors relating to the microscopic details of the  Langevin dynamics in a locally heated adiabatic potential.  

This behavior of the Fano factor can be rationalized based on the following observations. The only negative contributions to the integral over time in the current noise  (\ref{currentnoise}) are on the boundaries when the current crosses the mean. For the bi-stable case, this generally occurs only when the current switches between stable states. It is an intuitive notion to then conclude that larger switch rates will have an effect on decreasing the Fano factor (however having no switches at all will minimize it).

The size of the positive contribution to the current noise is dependent on two factors: firstly, the size of the fluctuations around the mean which correspond to the difference in current values between two configurations; and secondly, the ratio of time spent in each minimum. An increase to the applied voltage results in a larger fluctuation around the mean and as such, one would expect this to have an effect on increasing the Fano factor. However, this effect is counteracted by an increase to the mean current, which stays in the denominator of the Fano factor (\ref{fanofactor}). Additionally, the noise should be maximized when the mean current is directly in between our two current states; this occurs when the nucleus spends approximately equal time in each minimum. Therefore, the two  key parameters to control the Fano factor are the average switch rate (a single switch being a transition from one minimum to the other) shown in Fig.\ref{6c}, as well as the average waiting time (the average amount of time spent waiting in a minima before switching out) shown in Fig.\ref{6d}.  To maximize the Fano factor, one wants to keep the switch rate between conformations as small as possible but at the same the waiting times in both conformations should be comparable.  For example, let us consider the case of $V=0.05$.
As the voltage increases, the difference in effective temperatures  between the left and right minima increases as well, such that the left is substantially hotter, which will decrease the time spent waiting in the left minimum. In addition, the large applied voltage will physically deform the adiabatic potential in a manner akin to Fig.\ref{3b}, decreasing the depth of the left minimum relative to the right. These factors each act to decrease the left minimum waiting time relative to the right (see Fig.\ref{6d}). To compensate for this, the Fano factor peaks shifts towards smaller values of $k$, which act to deepen the left minimum, thus having the opposing effect of increasing the left minimum waiting time.

In Fig.\ref{FIG. 7}, we observe the Fano factor as the adiabatic potential transitions over the three possible regimes in our system as the spring constant $k$ is altered. The Fano factor demonstrates a strong dependency on instabilities within the system, undergoing a large peak as the bi-stable regime is entered, before decreasing back to sub-Poissonian values in the mono-stable regimes. The peak is shifted towards lower $k$ values for the reasons outlined regarding Fig\ref{6c}. The peak decreases slowly into the blue mono-stable regime because the stable minimum is very close to the left lead, which yields a small mean current in this region. As such our Fano factor according to (\ref{fanofactor}) is still large despite the adiabatic potential only being mono-stable.


\newpage

\section{conclusions}

In this paper, we studied current-induced atomic motion on molecule-electrode interfaces in molecular electronic junctions.
Structural changes on the interfaces are described in terms of a Langevin equation, which is obtained from the quantum mechanical first principles
in which we extract the slow nuclear dynamics from Wigner space Green's functions. The calculations of Green's functions and consequently all molecular junction observables include dynamical velocity-dependent corrections to include non-adiabatic effects of nuclear motion into the calculation of electronic properties.
We illustrate the theory by computing the transport properties of a model molecular junction: a single position-dependent resonant energy level which is coupled to the leads via a flexible (changing in time due to current flow) bond-length. The Langevin equation for the bond-length is integrated numerically and then the Green's functions, electric current, and current noise are computed along the stochastic trajectory.
We observe that even if the initial classical potential is harmonic, the effective adiabatic potential may develop bi-stability depending upon the parameters of the model. We mapped  the shapes of the adiabatic potential in the parameter space of the model. The different regimes for bistability depend critically on the interplay between the softness of the linking electrode-molecule bond and the coupling to the corresponding electrode.   

We introduce the concept of an effective local temperature using fluctuation-dissipation theorem ideas, which provides a useful insight on localized current-induced heating in molecular electronic junctions.  We observe a region of parameters in our junction  where the effective temperature becomes negative,  which means  the kinetic energy of nuclei will continue to increase if constrained to this region.  The structural instabilities and localized heating on molecule-electrode interfaces are quantified in terms of the current noise and Fano Factor.
 These demonstrated the influence of the calculated effective temperatures and adiabatic potentials on the nuclear dynamics, in which super-Poissonian Fano factors on the order of $\approx 400$ were observed.

\clearpage

\appendix

\section{Electronic diffusion coefficient}
\label{RWN}
\begin{widetext}
Computing the random white noise for our system starts by computing the quantity $\langle \delta f(t) \delta f(t')\rangle$. Noting that $\hat{f}(t) = f(t) + \delta f(t)$ allows us then to generate the expression

\begin{equation}
\langle \delta f(t) \delta f(t')\rangle = \langle \hat{f}(t) \hat{f}(t^{\prime}) \rangle - f(t) f(t^{\prime}),
\label{xyz}
\end{equation}
and so computing an expression for $\langle \delta f(t) \delta f(t')\rangle$ is reduced to calculating the quantities $\langle \hat{f}(t) \hat{f}(t^{\prime}) \rangle$ and $f(t) f(t^{\prime})$. 

The first term in (\ref{xyz}) term can be computed by making an explicit substitution for $f(t)$ in (\ref{yee}). This yields averages over strings of creation and annihilation operators which can be decomposed according to Wick's theorem. An example of which is given by

\begin{equation}
\langle a^{\dagger}_A a_B a^{\dagger}_C a_D \rangle = 
\
\langle a^{\dagger}_A a_B \rangle \langle  a^{\dagger}_C a_D \rangle 
\
+ \langle a^{\dagger}_A a_D \rangle \langle a_B a^{\dagger}_C \rangle ,
\end{equation}
where we have retained only the non-zero terms. It can be shown that the first term in these decompositions (involving no permutation of the creation/annihilation operators) will cancel exactly with the terms given in $f(t)f(t')$ (which can be easily calculated using (\ref{yeehat})). As a result, our random noise variance is then given by

\begin{multline}
\langle \delta f(t) \delta f(t') \rangle = 
\
\sum_{ij\bar{i}\bar{j}} \partial_x h_{ij} \mathcal{G}^>_{j\bar{i}}\partial_x h_{\bar{i}\bar{j}} \mathcal{G}^<_{\bar{j}i}
+ \sum_{ik\alpha \bar{i} \bar{k\alpha}} \Big(
\mathcal{G}_{i\bar{k\alpha}}^> \Lambda_{\bar{k\alpha} \bar{i}} \mathcal{G}_{\bar{i} k\alpha} \Lambda_{k\alpha i} 
+ \Lambda_{ik\alpha} \mathcal{G}^>_{k\alpha \bar{i}} \Lambda_{\bar{i}\bar{k\alpha}} \mathcal{G}_{\bar{k\alpha} i}
+ \mathcal{G}^>_{i\bar{i}} \Lambda_{\bar{i}\bar{k\alpha}} \mathcal{G}^<_{\bar{k\alpha}k\alpha} \Lambda_{k\alpha i}
\\
+ \Lambda_{ik\alpha}\mathcal{G}^>_{k\alpha \bar{k\alpha}} \Lambda_{\bar{k\alpha}\bar{i}} \mathcal{G}^<_{\bar{i}i} \Big)
+ \sum_{ij\bar{i}\bar{k\alpha}} \Big(\partial_x h_{ij} \mathcal{G}^>_{j\bar{k\alpha}} \Lambda_{\bar{k\alpha}\bar{i}} \mathcal{G}^<_{\bar{i}i}
+ \partial_x h_{ij} \mathcal{G}^>_{j\bar{i}} \Lambda_{\bar{i}\bar{k\alpha}} \mathcal{G}^<_{\bar{k\alpha}i} \Big)
+ \sum_{\bar{i}\bar{j}ik\alpha} \Big(\mathcal{G}^>_{i\bar{i}} \partial_x h_{\bar{i}\bar{j}} \mathcal{G}_{\bar{j}k\alpha} \Lambda_{k\alpha i}
+ \Lambda_{ik\alpha} \mathcal{G}^>_{k\alpha \bar{i}} \partial_x h_{\bar{i}\bar{j}} \mathcal{G}^<_{\bar{j}i} \Big),
\label{bigfella}
\end{multline}
where we have introduced our Green's functions. Here we use indices without a bar ($i$) to represent an operator acting at time $t$, while indices with a bar ($\bar{i}$) act at time $t'$. At this point, we must decompose our Green's functions into Green's functions in the system space and Green's functions in the leads. This involves applying our Dyson expansion to the Green's functions spanning the leads and system space (eg. $\mathcal{G}_{ik\alpha}$), as well as decomposing the $\mathcal{G}_{k\alpha \bar{k\alpha}}$ terms. For the purposes of this derivation, we will consider only a single term from (\ref{bigfella}) as the derivation can be applied similarly to the other terms in the equation. Consider

\begin{equation}
\sum_{ik\alpha \bar{i} \bar{k\alpha}} 
\mathcal{G}_{i\bar{k\alpha}}^> \Lambda_{\bar{k\alpha} \bar{i}} \mathcal{G}_{\bar{i} k\alpha} \Lambda_{k\alpha i}. 
\end{equation}

Applying our Dyson expansion to both Green's functions and taking advantage of the commutativity of matrix elements will yield

\begin{equation}
= \sum_{ik\alpha \bar{i} \bar{k\alpha}}
\
\Lambda_{\bar{k\alpha} \bar{i}} \Lambda_{k\alpha i}
\
\int  dt_1 \sum_m \Big(
\
\mathcal{G}^<_{\bar{i}m} v_{mk\alpha} \mathcal{G}^A_{k\alpha} + \mathcal{G}^R_{\bar{i}m} v_{mk\alpha} \mathcal{G}^<_{k\alpha} \Big)
\
\int  dt_2 \sum_n \Big(
\
\mathcal{G}^>_{in} v_{n\bar{k\alpha}} \mathcal{G}^A_{\bar{k\alpha}} + \mathcal{G}^R_{in} v_{n\bar{k\alpha}} \mathcal{G}^>_{\bar{k\alpha}} \Big).
\end{equation}

By expanding this product and introducing our self-energy like quantities, we find

\begin{equation}
= \int  dt_1 dt_2 \sum_{i\bar{i}mn} \Big(
\
\mathcal{G}^<_{\bar{i}m}(t',t_1) \Phi^A_{mi}(t_1,t)\mathcal{G}^>_{in}(t,t_2) \Phi^A_{n\bar{i}}(t_2,t')
\
+ \mathcal{G}^<_{\bar{i}m}(t',t_1) \Phi^A_{mi}(t_1,t)\mathcal{G}^R_{in}(t,t_2) \Phi^>_{n\bar{i}}(t_2,t') \Big).
\end{equation}

Applying a similar process for all terms in (\ref{bigfella}) yields the following expression:

\begin{multline}
\langle f(t) f(t') \rangle = 
\
\text{Tr} \Big\{ \partial_x h \mathcal{G}^> \partial_x h \mathcal{G}^<
\
+ \int  dt_1 dt_2 \Big(
\
\mathcal{G}^> \Phi^A \mathcal{G}^< \Phi^A + \mathcal{G}^R \Phi^> \mathcal{G}^< \Phi^A
\
\mathcal{G}^> \Phi^A \mathcal{G}^R \Phi^< + \mathcal{G}^R \Phi^> \mathcal{G}^R \Phi^<
\
+ \Psi^> \mathcal{G}^A \Psi^< \mathcal{G}^A \\ + \Psi^R \mathcal{G}^> \Psi^< \mathcal{G}^A
\
+ \Psi^> \mathcal{G}^A \Psi^R \mathcal{G}^< + \Psi^R \mathcal{G}^> \Psi^R \mathcal{G}^<
\
+ \mathcal{G}^> \Psi^< \mathcal{G}^A \Phi^A + \Psi^> \mathcal{G}^A \Phi^A \mathcal{G}^<
\
+ \mathcal{G}^> \Psi^R \mathcal{G}^< \Phi^A + \Psi^R \mathcal{G}^> \Phi^A \mathcal{G}^<
\\\
+ \mathcal{G}^> \Psi^R \mathcal{G}^R \Phi^< + \Psi^R \mathcal{G}^R \Phi^> \mathcal{G}^<
\
+ \mathcal{G}^> \Omega^< + \Omega^> \mathcal{G}^< \Big)
\\\
+ \int  dt_1 \Big(
\
d_x h \Big[\mathcal{G}^> \Phi^A \mathcal{G}^< + \mathcal{G}^R \Phi^> \mathcal{G}^< + \mathcal{G}^> \Psi^< \mathcal{G}^A + \mathcal{G}^> \Psi^R \mathcal{G}^<
\
+ \mathcal{G}^< \Phi^A \mathcal{G}^> + \mathcal{G}^R \Phi^< \mathcal{G}^> + \mathcal{G}^< \Psi^> \mathcal{G}^A + \mathcal{G}^< \Psi^R \mathcal{G}^> \Big] \Big) \Big\},
\label{bigD}
\end{multline}
where we have simplified the sum over central states into a trace and neglected time indices for brevity.

This equation must now be transformed into the Wigner space such that we can retrieve our diffusion coefficient. Beginning with 

\begin{equation}
\langle \delta f(t) \delta f(t') \rangle = D \delta (t-t'),
\end{equation}

We integrate both sides with respect to $\tau = t-t'$ which enables us to isolate $D$ as per

\begin{equation}
D =\int  d\tau \langle \delta f(t) \delta f(t') \rangle .
\end{equation}

In taking the Wigner transform, we once again consider a single example term which we will denote $D_1$.

\begin{align*}
D_1 &= \int  d\tau \int  dt_1 dt_2 
\
\mathcal{G}^> \Phi^A \mathcal{G}^< \Phi^A
\\
&=\int  d\tau A(t,t')B(t',t)
\\
&=\int  d\tau A(T,\tau)B(T,-\tau),
\end{align*}
where we have simply grouped terms together such that 
\begin{equation}
A(t,t') = \int dt_1 \mathcal{G}^> \Phi^A,
\end{equation}
and so on. Next, we transform it to the Wigner space using the Wigner convolution theorem and take the adiabatic limit, such that we obtain
\begin{align}
D_1  = \frac{1}{2\pi} \int  d\omega \widetilde{A}(T,\omega)\widetilde{B}(T,\omega).
\end{align}

All that then remains is to calculate the Wigner transform of our grouped variables $A$ and $B$ which is a relatively simple process. Applying this process to each term in \ref{bigD}, we find

\begin{multline}
D(x) = \frac{1}{2\pi} \int   d\omega \text{Tr}\Big\{ 
\partial_x h \widetilde{G}^> \partial_x h \widetilde{G}^< 
\
+ \widetilde{G}^> \widetilde{\Omega}^< + \widetilde{\Omega}^> \widetilde{G}^<
\
+ \widetilde{G}^> \widetilde{\Phi}^A \widetilde{G}^< \widetilde{\Phi}^A 
\
+ \widetilde{G}^R \widetilde{\Phi}^> \widetilde{G}^< \widetilde{\Phi}^A
\
+ \widetilde{G}^> \widetilde{\Phi}^A \widetilde{G}^R \widetilde{\Phi}^< 
\
+ \widetilde{G}^R \widetilde{\Phi}^> \widetilde{G}^R \widetilde{\Phi}^<
\\\
+ \widetilde{\Psi}^> \widetilde{G}^A \widetilde{\Psi}^< \widetilde{G}^A
\
+ \widetilde{\Psi}^R \widetilde{G}^> \widetilde{\Psi}^< \widetilde{G}^A
\
+ \widetilde{\Psi}^> \widetilde{G}^A \widetilde{\Psi}^R \widetilde{G}^<
\
+ \widetilde{\Psi}^R \widetilde{G}^> \widetilde{\Psi}^R \widetilde{G}^<
\
+ \widetilde{G}^> \widetilde{\Psi}^< \widetilde{G}^A \widetilde{\Phi}^A
\
+ \widetilde{G}^> \widetilde{\Psi}^A \widetilde{G}^A \widetilde{\Phi}^<
\\\
+ \widetilde{G}^> \widetilde{\Psi}^R \widetilde{G}^< \widetilde{\Phi}^A
\
+ \widetilde{\Psi}^R \widetilde{G}^> \widetilde{\Phi}^A \widetilde{G}^< 
\
+ \widetilde{\Psi}^> \widetilde{G}^R \widetilde{\Phi}^R \widetilde{G}^<
\
+ \widetilde{\Psi}^R \widetilde{G}^R \widetilde{\Phi}^> \widetilde{G}^<
\\\
+ d_x h\Big(\widetilde{G}^> \widetilde{\Phi}^A \widetilde{G}^< + \widetilde{G}^R \widetilde{\Phi}^> \widetilde{G}^< + \widetilde{G}^> \widetilde{\Psi}^< \widetilde{G}^A + \widetilde{G}^> \widetilde{\Psi}^R \widetilde{G}^< + \widetilde{G}^< \widetilde{\Psi}^A \widetilde{G}^> + \widetilde{G}^R \widetilde{\Phi}^< \widetilde{G}^> + \widetilde{G}^< \widetilde{\Psi}^> \widetilde{G}^A + \widetilde{G}^< \widetilde{\Psi}^R \widetilde{G}^> \Big) \Big\}.
\end{multline}

We observe that some terms are conjugates of each other, while certain strings of functions appear frequently in different terms. With significant simplification, we find our final expression for the diffusion coefficient as

\begin{multline}
D(x) = \frac{1}{2\pi} \int   d\omega \text{Tr} \Big\{
\
\partial_x h G^> \partial_x h G^< 
\
+ G^> \widetilde{\Omega}^< 
+ \widetilde{\Omega}^> G^< 
\\\
+ 2 \text{Re} \Big[ \Big(d_x h + \widetilde{\Psi}^R + \widetilde{\Phi}^A\Big)
\
\Big(G^< \widetilde{\Psi}^> G^A + G^> \widetilde{\Psi}^< G^A + G^> \widetilde{\Psi}^R G^< \Big)
\
+ \widetilde{\Psi}^> G^A \widetilde{\Psi}^< G^A + d_x h G^< \widetilde{\Psi}^R G^> \Big] \Big\}. 
\end{multline}

\end{widetext}

\newpage
%
 
\end{document}